\documentclass[11pt]{article}
\pdfoutput=1
\usepackage{jheppub}
\usepackage{subfigure,graphicx}

\newcommand\beq{\begin{equation}}
\newcommand\eeq{\end{equation}}
\newcommand\be{\begin{equation}}
\newcommand\ee{\end{equation}}
\def\be{\begin{eqnarray}}
\def\ee{\end{eqnarray}}

\def\Dslash{\,\,{\raise.15ex\hbox{/}\mkern-12mu D}}
\def\Dbarslash{\,\,{\raise.15ex\hbox{/}\mkern-12mu {\bar D}}}
\def\delslash{\,\,{\raise.15ex\hbox{/}\mkern-9mu \partial}}
\def\delbarslash{\,\,{\raise.15ex\hbox{/}\mkern-9mu {\bar\partial}}}
\def\pslash{\,\,{\raise.15ex\hbox{/}\mkern-9mu p}}
\def\calDslash{\,\,{\raise.15ex\hbox{/}\mkern-12mu {\cal D}}}

\newcommand{\Tr}{{\rm Tr}}

\def\lae{\mathrel{\mathop{\smash{\lower .5 ex \hbox{$\stackrel<\sim$}}}}}
\def\lae{\mathrel{\mathop{\smash{\lower .5 ex \hbox{$\stackrel>\sim$}}}}}

\title{Boundary Holographic Witten Diagrams}

\author[a]{Andreas Karch}
\author[a,b]{ and Yoshiki Sato}
\affiliation[a]{Department of Physics, University of Washington, Seattle, Wa, 98195-1560, USA}
\affiliation[b]{Department of Physics, Faculty of Science, The University of Tokyo, Bunkyo-ku, Tokyo 113-0033, Japan }

\preprint{\today}

\emailAdd{akarch@uw.edu,yoshiki3@uw.edu}

\abstract{In this paper we discuss geodesic Witten diagrams in generic holographic conformal field theories with boundary or defect.
Boundary CFTs allow two different decompositions of two-point functions into conformal blocks: boundary channel and ambient channel.
Building on earlier work, we derive a holographic dual of the boundary channel decomposition in terms of bulk-to-bulk propagators on lower dimensional AdS slices. In the situation in which we can treat the boundary or defect as a perturbation around pure AdS spacetime, we obtain the leading corrections to the two-point function both in boundary and ambient channel in terms of geodesic Witten diagrams which exactly reproduce the decomposition into corresponding conformal blocks on the field theory side.
}

\begin{document}
\maketitle


\section{Introduction}

Using the operator product expansion (OPE), correlation functions in conformal field theories naturally organize themselves into contributions from ``conformal blocks". These blocks sum up the contributions of all the descendants associated with a given primary operator arising in the OPE of a given pair of operators. In field theories with holographic dual, correlation functions in the bulk can be calculated from Witten diagrams \cite{Witten:1998qj}, that is position space Feynman diagrams in asymptotically anti-de Sitter (AdS) space. While the full correlation function must of course respect its decomposition into conformal blocks, the individual Witten diagrams do not nicely separate the contribution from any given block. It has been argued in \cite{Hijano:2015zsa} that one can isolate the contribution of a single conformal block to a conformal 4-pt function by calculating so called ``geodesic Witten diagrams". Geodesic Witten diagrams differ from their standard cousins in that the bulk interaction vertices are only integrated along a geodesic connecting two boundary operator insertions instead of over all of AdS space. One can show that these geodesic Witten diagrams represent the contribution of a single block by explicit calculation. But a more elegant method is to demonstrate that they obey the defining Casimir differential equations that blocks must obey together with the correct boundary conditions.

In conformal field theories with boundaries (bCFTs) the notion of conformal blocks becomes more interesting \cite{McAvity:1995zd}. The presence of a conformally invariant boundary reduces the conformal group in $d$ spacetime dimensions from $SO(d,2)$ to $SO(d-1,2)$. The reduced symmetry allows for the appearance of a non-trivial function depending on a single conformally invariant cross-ratio already at the level of the 2-pt function. This correlator can be decomposed into conformal blocks in two distinct ways: in the ``ambient space channel" one uses the standard ambient space\footnote{As in \cite{Aharony:2003qf} we use the term ambient space for the $d$-dimensional space-time (labelled by indices $\mu$, $\nu$) in which the $d-1$ dimensional defect (labelled by $i$, $j$) is embedded. The direction transverse to the defect is called $w$. We reserve the term ``bulk" for the $d+1$ dimensional spacetime of the holographic dual given by \eqref{metric}.} 
OPE to re-express the two-point function as a sum over one-point functions (which need not vanish in a bCFT). The contribution of a given primary and its descendants gets summed up into an ``ambient block". In the ``boundary channel" one uses a novel operators expansion, the BOPE or boundary operator expansion, to expand any ambient operators in terms of boundary localized operators \cite{McAvity:1995zd}. 
This way the full ambient space 2-pt functions gets reduced to a sum over 2-pt functions of boundary localized operators. Once again, the contribution of a given primary and its descendants can be summed up into a ``boundary block". Demanding equality of the decompositions into ambient blocks and boundary blocks gives interesting constraints on the bCFT data, as encapsulated in the boundary bootstrap program \cite{Liendo:2012hy}.

It is natural to expect that the conformal block decompositions of the 2-pt function in a bCFT with holographic dual can once again be captured by geodesic Witten diagrams. First steps in this direction have been taken in \cite{Rastelli:2017ecj}. bCFTs are dual to a spacetime with $d+1$ non-compact directions that allow a slicing in terms of AdS$_d$ \cite{Karch:2000ct}. The simplest models have a $d+1$ dimensional bulk given by a metric
\beq
\label{metric}
\mathrm{d}s^2 = \mathrm{e}^{2 A(r)} \mathrm{d}s^2_{\mathrm{AdS}_d} + \mathrm{d}r^2
\eeq
potentially times some internal space. If $\mathrm{e}^A = \cosh(r/L)$ (hereafter we take the AdS radius $L=1$) this metric is simply AdS$_{d+1}$. For the holographic dual of  genuine bCFTs the standard holographic dictionary requires the warpfactor $A$ to approach this asymptotic form for large $r$. Examples in this class include AdS sliced Randall-Sundrum models \cite{Karch:2000ct,Karch:2000gx}, the very closely related AdS/bCFT proposal by Takayanagi \cite{Takayanagi:2011zk}, as well as the $d=4$ Janus solution of type IIB supergravity \cite{Bak:2003jk} together with its cousins in other $d$. The former two are toy models, based on Einstein gravity coupled to branes with tension. They have no known embedding in string theory and no explicitly known dual field theory. The latter is an explicit top-down solution; its dual field theory consists of ${\cal N}=4$ super Yang-Mills (SYM) theory with a step function defect across which the coupling constant jumps\footnote{Solutions like the Janus solution in which ambient space extends on both sides of the defect with different properties are often referred to as (holographic duals of) interface conformal field theories, or iCFTs. If the field theory on both sides of the interface is the same but extra degrees of freedom are localized on it, the system is often referred to as a defect CFT or dCFT. Both iCFTs and dCFTs can be seen as special cases of bCFT by employing the folding trick: the interface/defect can be viewed as a boundary in a theory whose ambient space contains two decoupled copies of the original CFT on half-space with interactions localized on the boundary.}. The non-trivial metric \eqref{metric} is supported by matter fields whose profile is independent of the coordinates on the slice. We'll collectively denote these background fields as $X(r)$. For example, in the Janus solution there is a single scalar turned on (the dilaton) with an $r$-dependent profile. In more general holographic bCFT constructions, such as the dual of ${\cal N}=4$ SYM on half-space with supersymmetry preserving boundary conditions \cite{DHoker:2007zhm,DHoker:2007hhe,Aharony:2011yc}, the warpfactor $A$ depends non-trivially also on the compact internal space. In this case we can still use the metric \eqref{metric} with the understanding that in this $D$ dimensional metric $r$ stands for the set of all internal variables, $\mathrm{d}r^2$ is the $D-d$ dimensional metric on the internal space, and $A(r)$ really is, in general, a function of all these $D-d$ internal coordinates.

In \cite{Rastelli:2017ecj} only the simplest case of a holographic dCFT was addressed. In the case where the bCFT is really a dCFT with a small number of matter fields localized on a defect in a large $N$ gauge theory, one can neglect the backreaction of the matter fields on the ambient space field theory. A simple top-down example of such a ``probe brane dCFT" is the D3/D5 system of \cite{Karch:2000gx,DeWolfe:2001pq}, representing ${\cal N}=4$ SYM coupled to a $2+1$ dimensional hypermultiplet in the fundamental representation of the gauge group. In this case, the dual geometry remains AdS$_5$ throughout (that is $\mathrm{e}^A=\cosh(r)$ for all values of $r$). The defect is dual to a probe D5 brane living on one of the AdS$_4$ slices\footnote{For the simplest case of a D5 brane intersecting $N$ D3 branes the probe is located at $r=0$, but it can move to a different $r_*$ if we let some of the D3 branes end on the D5 brane \cite{Karch:2000gx}.}. 
In this case the prescription for geodesic Witten diagrams is fairly straightforward as both geodesic and propagator retain their standard AdS form and all one has to account for are the extra brane localized matter and interactions. The resulting proposals of \cite{Rastelli:2017ecj} can once again be confirmed by explicit calculation as well as by the Casimir method. But the prescriptions as phrased in this work heavily rely on the special probe brane scenario and it remained far from clear of how to implement the idea of geodesic Witten diagrams in generic boundary conformal field theories. It is our aim in this work to fill this gap.

In fact, the bulk manifestation of boundary conformal blocks has been understood in a slightly different context before. In \cite{Aharony:2003qf} it was shown that the BOPE manifests itself as a mode decomposition in the bulk. This construction was used in \cite{Karch:2017fuh} to show that particular integrals of bulk scalar fields along geodesics, the so called weighted X-ray transforms, are the correct bulk duals to the boundary conformal blocks. They naturally live in ``boundary kinematic space". We will review both these constructions in the next section, as they will play a crucial role in deriving the correct geodesic Witten diagram prescriptions for generic holographic bCFTs. In section \ref{sec3} we will give this derivation of the diagrams associated to the contribution of a single conformal block in the boundary and ambient channel respectively. After presenting some simple examples in section \ref{sec4} we will conclude in section \ref{sec5}.

\section{Holographic Boundary Operator Expansion and Boundary Blocks}

Underlying the decomposition of the 2-pt function into boundary conformal blocks is the notion of a boundary operator expansion (or BOPE). As was demonstrated in \cite{Aharony:2003qf}, the BOPE in the bulk can naturally be understood as a mode-decomposition
of fields living on an AdS$_d$ sliced geometry with metric \eqref{metric}. We want to solve the equation of motion for a scalar field $\phi_{d+1}(r,y)$, dual to an ambient space operator $O$ of dimension $\Delta$. Here $y$ and $r$ stand collectively for the coordinates along the AdS$_d$ slice and the transverse directions respectively. Let us for concreteness focus on the case with only on internal variable $r$; the generalization to many $r$ is straight forward as we will see as we go along. Using the $d+1$ dimensional geometry \eqref{metric} with background fields $X(r)$ turned on the equation of motion reads
\beq
\label{eom}
(\Box _g - M^2(r)) \phi_{d+1} = ( {\cal D}^2_r + \mathrm{e}^{-2 A} \partial_d^2 - M^2(r) ) \phi_{d+1} = 0 \,. \eeq
$\Box_g$ is the Laplacian in the full Janus background geometry \eqref{metric} and
$\partial_d^2$ stands for the AdS$_d$ Laplacian on the slice. The radial operator ${\cal D}^2_r$ is defined as
\beq
{\cal D}^2_r \psi _n(r) \equiv
\psi''_n(r) + d A'(r) \psi'_n(r) \,.
\eeq
The potential term $M^2(r)$ includes the bulk scalar mass $M_0^2$, but also all the interactions with the background fields $X(r)$\footnote{The dilaton is special because its coupling appears in front of a kinetic term. We will treat this case in section \ref{sec4} as an example of our prescription.}. For example a quartic $X^2 \phi^2$ coupling in the Lagrangian will give rise to an extra $X^2(r)$ term in $M^2$. The only important property we need from $M$ is that it does not depend on the $y$ coordinates as guaranteed by the $SO(d-1,2)$ defect conformal symmetry.
We can make a separation of variables ansatz
\beq
\label{sov}
\phi_{d+1} (r,y) = \sum_n \psi_n(r) \phi_{d,n} (y)
\eeq
so modes $\phi_{d,n}$ obey a standard scalar wave equation on the slice
\beq \label{onslicefree} \partial_d^2 \phi_{d,n} = m_n^2 \phi_{d,n} \,. \eeq
The eigenvalues $m_n^2$ are then determined by the internal mode equation:
\beq
\label{modeeq}
{\cal D}^2_r \psi _n(r)  - M^2(r) \psi_n(r) = - \, \, \mathrm{e}^{-2 A(r)} m_n^2 \psi_n(r) \, .
\eeq
This 2nd order differential equation can easily be brought into the form of a 1d Schr\"odinger equation by a simple change of variables\footnote{\label{foot5} A change of variables from $r$ to a conformal coordinate $z$ with $\mathrm{d}r=\mathrm{e}^A \mathrm{d}z$ removes
the $\mathrm{e}^{-2A}$ factor in front of the eigenvalues $m_n^2$ and a further rescaling $\psi_n = \mathrm{e}^{-(d-1)A/2} \Psi_n$ eliminates
the first derivatives acting on the mode-function so that we are left with a standard Schr\"odinger equation for $\Psi_n(z)$ together with its usual norm and an effective potential of \cite{DeWolfe:1999cp}
\beq
\label{analog}
V(z) = \frac{1}{2} \left [ \left ( \frac{d-1}{2} \frac{\mathrm{d}A}{\mathrm{d}z} \right )^2 + \frac{d-1}{2} \frac{\mathrm{d}^2A}{\mathrm{d}z^2} + M^2 \mathrm{e}^{2A} \right ]\, .
\eeq
}; correspondingly the modefunctions can be chosen to be complete and orthonormal with respect to the Schr\"odinger norm, which in the original variables implies
\beq
\label{orthocomplete}
\sum_n \psi_n(r) \psi_n(r') = \mathrm{e}^{-(d-2)A(r)} \delta(r-r') \, , \qquad  \int \! \mathrm{d}r \, \mathrm{e}^{(d-2) A(r)} \, \psi_m \psi_n = \delta_{mn}\,, \eeq
and the eigenvalues $m_n^2$ are real.
For cases with more than one internal coordinate or couplings of $X(r)$ to terms involving derivatives of $\phi_{d+1}$, it is straightforward to write down the corresponding eigenvalue problem. This is the only ingredient that needs to be changed in these cases. 

Since each mode $\phi_{d,n}$ on the slice obeys a scalar wave equation on AdS$_d$, it is dual to an $SO(d-1,2)$ primary operator $o_{n}$ localized on the defect. It was shown in \cite{Aharony:2003qf} that these $d-1$ dimensional operators are exactly the
ones appearing in the BOPE of $O$, that is they appear in the expansion of $O$ in terms of boundary localized operators
\beq
\label{bope}
O(\vec{x},w) = \frac{c_{\mathbf{1}}^O}{(2w)^\Delta} +\sum_k \frac{c^O_{k}}{(2w)^{\Delta - \Delta_k}} o_{k}(\vec{x}) \,.
\eeq
We broke out the contribution $c_{\mathbf{1}}^O$ of the identity operator for clarity.
In the BOPE as written in \eqref{bope} the operators on the right hand side, labeled by $k$, are both primaries and descendants. However the contribution of the descendants are completely determined by that of the primaries, labeled by $n$, and they can be summed into non-local block operators ${\cal B}_n(\vec{x},w)$ so that the BOPE reads:
\beq
O(\vec{x},w) = \frac{c_{\mathbf{1}}^O}{(2w)^\Delta} +\sum_n c^O_{n} {\cal B}_n(\vec{x},w) \,.
\eeq
The functional form of the blocks is uniquely fixed by symmetry \cite{McAvity:1995zd}.
According to \cite{Aharony:2003qf} the primaries appearing in the BOPE are in 1-to-1 correspondence with the modes $\phi_{d,n}$, their dimensions are given by the eigenvalues $m_n^2$ by the usual AdS$_d$ relation
\beq
\Delta_n [ \Delta_n-(d-1) ] = m_n^2
\eeq
and the OPE coefficients $c^O_{n}$ are encoded in the asymptotic fall-off of the modefunctions $\psi_n$ \cite{Aharony:2003qf}.

The decomposition of an ambient operator $O$ into conformal blocks can be inserted into any correlation function which reduces the ambient space correlator into correlators of the non-local blocks. In order to implement this procedure for the special case of the 2-pt function\footnote{In the literature the name ``conformal block" is often used both for the non-local operator ${\cal B}$ itself, but also for the non-trivial function of cross-ratios it contributes to a particular correlation function (usually the 4-pt function in CFTs without boundary and the 2-pt function in bCFTs). We will try to be careful in the following to reserve the name for the operator itself and will refer to the non-trivial function of the cross-ratio appearing in the 2-pt function as the contribution from a particular block.}, we would like to get a bulk representation of the conformal block itself. This was provided in \cite{Karch:2017fuh} using the construction of boundary kinematic space, which generalized the previously introduced kinematic space of \cite{Czech:2015qta,Czech:2016xec,deBoer:2016pqk,daCunha:2016crm} to bCFTs. It was shown that from the bulk field $\phi_{d+1}$ one can construct kinematic space operators $R_n \phi$ by a weighted X-ray transform
\beq R_n \phi(y) = \int_{\gamma} \! \mathrm{d}r \,  \mathrm{e}^{(d-2) A}\psi_n(r) \phi_{d+1}(r,y)\,. \eeq
Here $\gamma$ is a symmetry enhanced geodesic, as defined in \cite{Karch:2017fuh}, emanating from the boundary point $y$ at $r\rightarrow \infty$. Of course there are many geodesics anchored at this point and their detailed properties depend on the warp-factor $A$. But it was shown in \cite{Karch:2017fuh} that the line
\beq
y(r) = \text{const.}
\eeq
is a geodesic for any choice of warpfactor $A(r)$ and is in fact singled out to be the only one compatible with the expected symmetries of the block. This is the geodesic $\gamma$ that appears in the weighted X-ray transform. The weights $\psi_n$ are exactly the mode-functions appearing in \eqref{modeeq}. Using the orthogonality of the modefunctions as well as the
mode decomposition \eqref{sov} we see that the weighted X-ray transform exactly pulls out the AdS$_d$ modes $\phi_{d,n}$:
\beq
\label{xray}
R_n \phi(y) = \phi_{d,n}(y) \, .
\eeq
Writing the AdS$_d$ metric on the slice parametrized by the $d$ coordinates $y$ as\footnote{The coordinate $w$ transverse to the defect also corresponds to a radial direction on the AdS$_{d}$ slice.}
\beq
\label{ppatch}
\mathrm{d}s^2_{\mathrm{AdS}_d} = \frac{\mathrm{d}w^2 + \mathrm{d}\vec{x}^2}{w^2}
\eeq
it was further shown in \cite{Karch:2017fuh} that $R_n \phi(\vec{x},w)$ was in fact equal to the conformal block ${\cal B}_n(\vec{x},w)$, that is the radial direction along the slice takes the role of the direction orthogonal to the defect. In the following section we will use this insight to derive the contribution of a given boundary conformal block to the 2-pt function of two ambient space operators in terms of geodesic Witten diagrams.

\section{Block decomposition of the 2-pt function}
\label{sec3}

\subsection{2-pt function in bCFT}
\subsubsection{General Structure}

In a conformal field theory without boundary, the conformal block decomposition is usually applied to 4-pt functions since they are the simplest correlators that allow any non-trivial functional dependence on the position of the insertion points. With 4 insertion points one can form 2 conformally invariant cross-ratios and the correlator can be a non-trivial function of both of them which can be decomposed (in different ways) into contributions from blocks appearing in the OPE applied to two operators at a time. For the case of a bCFT, a non-trivial cross-ratio already appears in the 2-pt function $\langle O_{1}(\vec{x}_1,w_1)
O_{2} (\vec{x}_2,w_2) \rangle$ of two ambient space primary operators.
\beq \eta=\frac{(\vec{x}_1 - \vec{x}_2)^2 + (w_1 - w_2)^2 }{w_1 w_2} \eeq
is conformally invariant and correspondingly the general form of the 2-pt function is
\beq
\label{form}
 \langle O_{1}(\vec{x}_1,w_1)
O_{2} (\vec{x}_2,w_2) \rangle = \frac{f(\eta)}{(2 w_1)^{\Delta_1} (2 w_2 )^{\Delta_2}} \,. \eeq
Note that, unlike in the case without boundary, this 2pt-function need not vanish in the case when the two operators have different dimensions $\Delta_1$ and $\Delta_2$.

Both ambient and boundary block expansions can in principle be applied to the general case of two different operators. In the following we will however restrict ourselves to the case where both insertions are the same operator $O$ with dimension $\Delta$. Applying the BOPE to $O(\vec{x}_1,w_1)$ and $O(\vec{x}_2,w_2)$ separately,
the 2-pt function can be written as a sum of contributions from 2-pt functions of the blocks. The residual boundary conformal group insures that the only non-vanishing 2-pt functions arise from one and the same primary $o_{n}$ appearing in both BOPEs. Furthermore, for 2-pt functions of scalar operators angular momentum conservation implies that only blocks build from scalar operators can contribute. This allows one to give a boundary-channel
expansion for $f(\eta)$ of the form
\begin{equation}
\label{boundaryf}
f(\eta) = (c^{O}_{\mathbf{1}})^2 + \sum_n (c^O_n)^2 f_{\partial} (\Delta_n,\eta) \,.
\end{equation}
$c^{O}_{\mathbf{1}}$ denotes the contribution of the identity operator.
The contribution of the $n$-th boundary block, $f_{\partial}(\Delta_n,\eta)$, is fixed by conformal invariance\footnote{As remarked before, sometimes $f_{\partial}$ itself is referred to as the block, but for clarity we reserve this name for the non-local operator appearing the BOPE, not its contribution to the 2-pt function.}. The explicit form of $f_{\partial}$ can either be obtained by summing up the contributions of the descendants \cite{McAvity:1995zd} or, more elegantly, by the Casimir method \cite{Liendo:2012hy}: since all the descendants of a given primary sit in the same representation of the conformal algebra, they all have to correspond to the same eigenvalue of the Casimir operator $L_{\partial}^2$ of the conformal group. In terms of the generators $D$ (dilatation), $K_i$ (special conformal), $P_i$ (translations), and $M_{ij}$ (boosts and rotations) in their standard representation as differential operators acting on a scalar at $(\vec{x},w)$ one has
\beq L_{\partial}(\vec{x},w)^2 = - 2 D^2 - (K^i P_i + P^i K_i) + M^{ij} M_{ij} \eeq
with
\begin{equation}
\begin{split}
D=i (w \partial_w + x^i \partial_i)\,, &\qquad  K_i =  i (2x_i (x^j \partial_j + w \partial_w) - (x^j x_j + w^2) \partial_i)\,, \\
P_i = - i \partial_i\,, &\qquad  M_{ij} = - i (x_i \partial_j - x_j \partial_i) \, .
\end{split}
 \end{equation}
The differential equation
\beq  (L_{\partial}(\vec{x}_1,w_1) + L_{\partial}(\vec{x}_2,w_2) )^2 \langle O(\vec{x}_1,w_1)
O(\vec{x}_2,w_2) \rangle = \Delta_n (\Delta_n-d +1) \langle O(\vec{x}_1,w_1)
O(\vec{x}_2,w_2) \rangle \eeq
together with the boundary condition that as the ambient operators approach the defect ($\eta \rightarrow \infty$) the contribution of the blocks is dominated by the primary or in other words
\beq
f_\partial(\Delta_n, \eta) \sim \eta^{-\Delta_n}
\eeq
gives the contribution of the $n$-th block as \cite{Liendo:2012hy}
\beq
\label{boundaryblock}
f_{\partial} (\Delta_n,  \eta)= \left ( \frac{\eta}{4} \right )^{- \Delta_n} {}_2 F_1 \left(\Delta_n,
\Delta_n - \frac{d}{2} +1,2 \Delta_n  -d+ 2,- \frac{4}{\eta} \right ) \, .
\eeq
A different expansion of $f(\eta)$ can be obtained by using the standard ambient space OPE on the product $O(\vec{x}_1,w_1) O(\vec{x}_2,w_2)$. This reduces the calculation of the 2-pt function to a sum over 1-pt functions of all the operators appearing in the OPE of $O$ with itself. In a standard CFT the only non-vanishing 1-pt function comes from the identity operator. So the 2-pt function is completely determined by the identity block. In a bCFT any scalar primary operator can have a non-trivial 1-pt function
\beq
\label{vev}
\langle O(\vec{x},w) \rangle = \frac{a_O}{(2 w)^{\Delta}}
\eeq
and so the 2-pt function of ambient operators can be reduced to a sum over 1-pt functions. This give us an expansion of $f$ of the form
\beq
\label{bulkf}
f(\eta) = \lambda_{\mathbf{1}} \left ( \frac{4}{\eta} \right )^{\Delta} + \sum_N \lambda_N a_N f_{\text{ambient}} (\Delta_N,\eta)
\eeq
where the sum over $N$ is a sum over ambient space primaries (that is primaries under the full $SO(d,2)$), $\lambda_N$ are the OPE coefficients and $a_N$ the constants determining their 1-pt functions according to \eqref{vev}. The contribution of the $N$-th ambient block can once more be obtained by explicit summation or the Casimir method \cite{Liendo:2012hy}:
\beq
f_{\text{ambient}}(\eta) = \left ( \frac{\eta}{4} \right )^{\Delta_N/2 -  \Delta} {}_2 F_1 \left(\frac{\Delta_N}{2},\frac{\Delta_N}{2},\Delta_N-\frac{d}{2} +1 , - \frac{\eta}{4} \right) \,.
\eeq
Like $f_{\partial}$, $f_{\mathrm{ambient}}$ can be found as an eigenfunction of a conformal Casimir operator, but this time it is the full $SO(d,2)$ Casimir (acting on $\vec{x}_1$, $w_1$, $\vec{x}_2$, $w_2$) and the eigenvalue is $\Delta_N (\Delta_N-d)$. Equating the two expansion gives rise to the boundary bootstrap equation \cite{Liendo:2012hy}, which will however not play a major role in this work.

\subsubsection{Simple Examples: Dirichlet, Neumann and ``No-brane"}
\label{examples}

The simplest examples to illustrate the various block decompositions are free field theories. The three cases one wants to distinguish are a free scalar bCFT with Dirichlet boundary conditions (the ``Dirichlet theory"), a free scalar bCFT with Neumann boundary conditions (the ``Neumann theory") or the ``free No-braner" theory: a free scalar $\varphi$ without boundary or interface in which one randomly picks the $w=0$ surface to be treated as an interface. As emphasized in \cite{Aharony:2003qf}, in a free theory the BOPE is essentially a Taylor expansion. Correspondingly, the defect operators $o_n$ are built from $O(\vec{x},w=0)$ and $\partial_w O(\vec{x},w=0)$ and so on\footnote{As shown in \cite{Aharony:2003qf} these $w$ derivatives of $O$ are not primaries by themselves, but one can build primaries from linear combination of $w$ derivatives and operators built form derivatives along the slice.}. Due to the equations of motion, in a free theory $\partial_w^2 O$ for $O=\varphi$ is related to the on-slice Laplacian and hence is already a descendent of the defect conformal algebra. Correspondingly, the only operators that can appear in the BOPE are $O$ (dimension $\Delta_{\varphi}=d/2-1$) and $\partial_w O$ (with dimension $d/2$). The Dirichlet theory only has the former, the Neumann theory the latter and the free no-braner has both. Correspondingly the boundary channel expansion yields \cite{Liendo:2012hy}
\begin{align}
f_{\text{Dirichlet}}(\eta) &= \left ( \frac{4}{\eta} \right )^{\Delta_{\varphi}} \left [ 1 - \left ( \frac{\eta}{\eta+4}
\right )^{\Delta_{\varphi}} \right ] \\
f_{\text{Neumann}}(\eta) &= \left ( \frac{4}{\eta} \right )^{\Delta_{\varphi}} \left [ 1 + \left ( \frac{\eta}{\eta+4}
\right )^{\Delta_{\varphi}} \right ] \\
f_{\text{no-brane}}(\eta) &= \frac{1}{2} \left ( \frac{4}{\eta} \right )^{\Delta_{\varphi}} \left [ 1 + \left ( \frac{\eta}{\eta+4}
\right )^{\Delta_{\varphi}} +1 - \left ( \frac{\eta}{\eta+4}
\right )^{\Delta_{\varphi}}\right ] = \left ( \frac{4}{\eta} \right )^{\Delta_{\varphi}}
\end{align}
which means that the full correlator takes the expected form one would get from a method of images construction
\beq \langle \varphi(\vec{x}_1,w_1) \varphi (\vec{x}_2,w_2) \rangle  =
\left \{ \begin{array}{ll} \frac{1}{(\Delta x^2)^{\Delta_{\varphi}}}  - \frac{1}{(\Delta x_{\mathrm{R}}^2)^{ \Delta_{\varphi}}} & \mbox{ for Dirichlet} \\ \frac{1}{(\Delta x^2)^{ \Delta_{\varphi}}}
+ \frac{1}{(\Delta x_\mathrm{R}^2)^{\Delta_{\varphi}}} & \mbox{ for Neumann} \\
\frac{1}{(\Delta x^2)^{\Delta_{\varphi}}} & \mbox{ for no-brane} \end{array} \right .
\eeq
where
\beq
\Delta x^2 = (\vec{x}_1 - \vec{x}_2)^2 + (w_1 - w_2)^2 \, , \qquad
\Delta x_{\mathrm{R}}^2 = (\vec{x}_1 - \vec{x}_2)^2 + (w_1 + w_2)^2 \, .
\eeq
Instead of interpreting the mirror charge terms as the contribution of boundary blocks with dimension $\Delta_{\varphi} +1$ and $\Delta_{\varphi}$ in the Dirichlet and Neumann case respectively, we can also give them an ambient channel representation: they are the ambient block of dimension $2 \Delta_{\varphi}$ associated with the operator $\varphi^2$ appearing in the operator product of $\varphi$ with itself. The difference in sign comes from the difference in vacuum expectation value of $\varphi^2$.

Note that the ``no-braner" construction can be applied to any CFT, be it interacting or not. The fact that the BOPE truncates to only two primaries is special to the case of a free no-braner. In particularly, in the ``no-braner" of a theory with holographic duals, one finds the entire tower of fields with dimension $\Delta_n = \Delta +n$ associates to the primaries built from $w$-derivatives of $O$ in the BOPE of $O$ \cite{Aharony:2003qf}.

\subsubsection{Holographic calculation}

\begin{figure}
        \centering
        \includegraphics[scale=0.3]{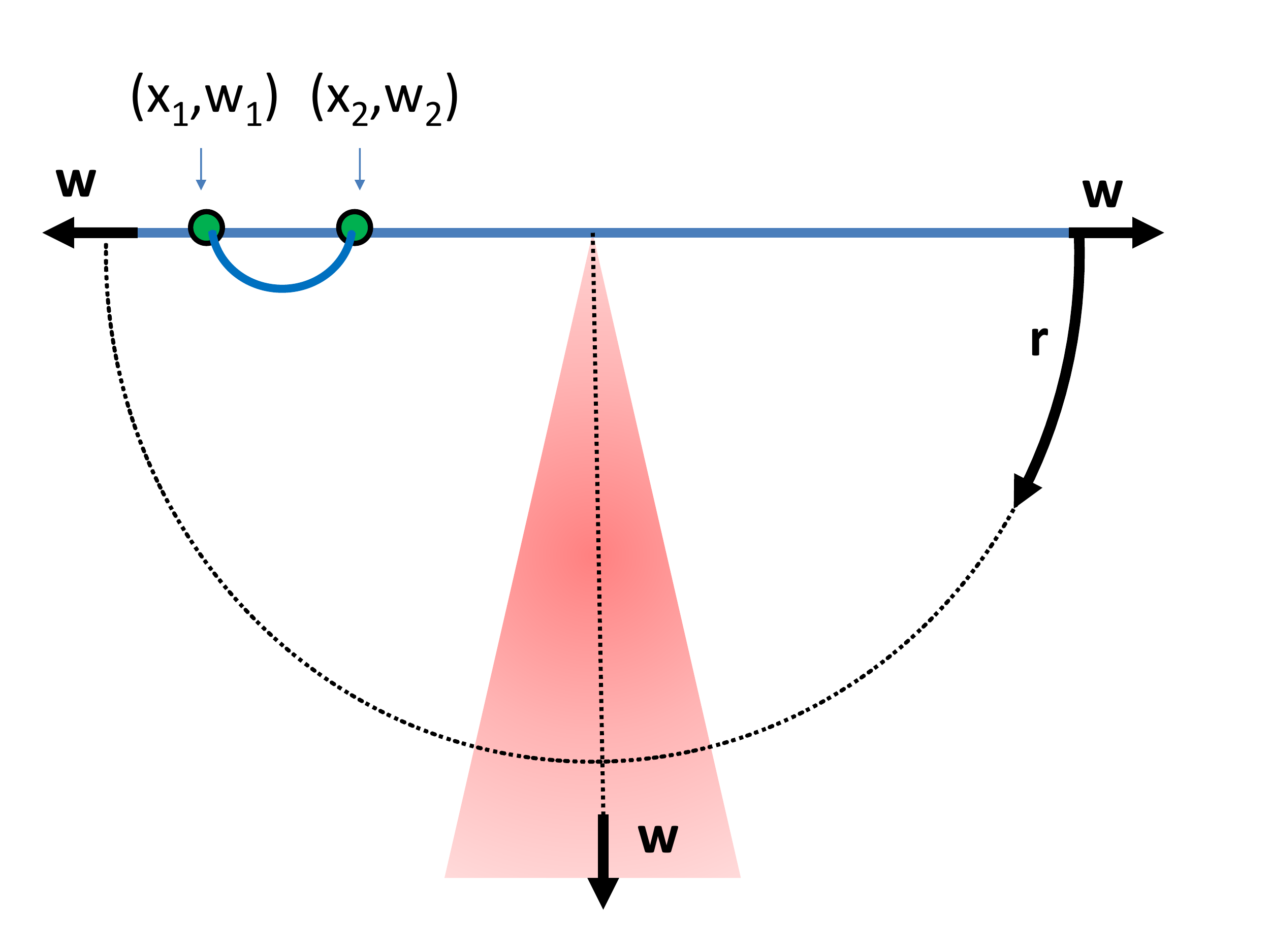}
		\caption{Full 2-pt function in an iCFT. The red triangle indicates the region in which the metric strongly deviates from AdS$_{d+1}$. In a bCFT space would truncate smoothly inside the triangle.
We also indicated how the $r$ and $w$ coordinates used in the main text parametrize the space depicted in this figure. All following figures also depict the same $r$-$w$ plane.} \label{fig:full}
\end{figure}

In principle it is easy to calculate a dCFT 2-pt function from Witten diagrams: we simply need to obtain the bulk-to-boundary propagator in the full holographic dCFT geometry of \eqref{metric}. This amounts to calculating a single Witten diagram as depicted in figure \ref{fig:full}. So to some extend there is much less urgency in this case to organize the contributions according to conformal blocks. Nevertheless, it may sometimes be convenient to do so. We can use the exact representation in terms of the diagram of figure \ref{fig:full} in order to derive the geodesic Witten diagrams associated with the block decomposition.

In principle the Witten diagram in figure \ref{fig:full} is very easy to calculate. In terms of the bulk-to-boundary propagator
$K_{\Delta,d+1}(r,\vec{x}_1,w_1,\vec{x}_2,w_2)$ we have\footnote{For the propagators and their relation to the correlation functions we are following the conventions of \cite{Ammon:2015wua}.}
\beq
\langle O(\vec{x}_1,w_1) O(\vec{x}_2,w_2) \rangle = C \lim_{r \rightarrow \infty} \mathrm{e}^{- \Delta r}
K_{\Delta,d+1}(r,\vec{x}_1,w_1,\vec{x}_2,w_2) \,.
\eeq
The constant $C$ encodes the prefactor of the action\footnote{For a scalar field,
we take the action to be $$S=-\frac{C}{2} \int \! \mathrm{d}^{d+1} x \, \sqrt{-g}
(g^{MN} \partial_M \phi_{d+1} \partial_N \phi_{d+1} + M^2(r) \phi_{d+1}^2)\,,$$
where $M,N$ label $d+1$ coordinates of the bulk spacetime.
}. The bulk-to-boundary propagator is,
as usual, defined as a solution to the scalar equations of motion
\beq (\Box_g - M^2(r)) K_{\Delta,d+1} =0
\eeq
approaching the appropriate delta function at the boundary
\beq
\lim_{r \rightarrow \infty} \mathrm{e}^{(d-\Delta)r} K_{\Delta,d+1}(r,\vec{x}_1,w_1,\vec{x}_2,w_2) = \delta(\vec{x}_1 - \vec{x}_2) \delta(w_1 - w_2) \,.
\eeq
As in \eqref{modeeq}, $M^2(r)$ is a non-trivial function of the radial direction not just encoding the mass of the scalar field but also its interactions with all the matter fields that have a non-trivial profile in the background geometry. Asymptotically the metric approaches that of AdS$_{d+1}$ and all matter fields go to constants so that $M^2(r)$ approaches $M_0^2$ which is related to the dimension $\Delta$ by the usual
\beq
\Delta (\Delta-d) = M_0^2 \,.
\eeq
The bulk-to-boundary propagator can also be obtained as a limit of the full bulk-to-bulk propagator $G_{\Delta,d+1}$ which obeys
\beq
\label{propeq}
\sqrt{-g} (\Box_g - M^2(r_1) ) G_{\Delta,d+1}(r_1,\vec{x}_1,w_1,r_2,\vec{x}_2,w_2) = \delta(r_1-r_2) \delta(\vec{x}_1 - \vec{x}_2) \delta(w_1-w_2) \,.
\eeq
We can recover $K_{\Delta,d+1}$ via\footnote{As one approaches the boundary of asymptotically AdS$_{d+1}$ space, the metric diverges as $f^{-2}$ where the ``defining function" $f$ has a single zero at the boundary. To extract the asymptotic behavior of the various fields on this space, one multiplies with the appropriate powers of $f$. In the metric \eqref{metric} one naively may have expected to use $\cosh(r) \sim \mathrm{e}^r/2$ as the defining function and hence multiply $K_{\Delta,d+1}$ simply with $\mathrm{e}^{\Delta r}/2^{\Delta}$. However in this case one would obtain answers relevant for a field theory on AdS$_d$. If we are interested in extracting correlators for a flat space bCFT, we need to use $f=\mathrm{e}^r/(2w)$ as the defining function.}
\beq
\label{bulktobound}
K_{\Delta,d+1}(r_1,\vec{x}_1,w_1,\vec{x}_2,w_2) = \lim_{r_2 \rightarrow \infty} (2 \Delta-d) \frac{\mathrm{e}^{ \Delta r_2}}{(2w_2)^{\Delta}}
G_{\Delta,d+1}(r_1,\vec{x}_1,w_1,r_2,\vec{x}_2,w_2)
\eeq
so that
\beq
\label{twopoint}
\langle O(\vec{x}_1,w_1) O(\vec{x}_2,w_2) \rangle = C (2 \Delta-d)^2 \lim_{r_{1,2} \rightarrow \infty} \frac{\mathrm{e}^{ \Delta (r_1 + r_2)}}{(2w_1)^{\Delta} (2w_2)^{\Delta}}
G_{\Delta,d+1}(r_1,\vec{x}_1,w_1,r_2,\vec{x}_2,w_2) \,.
\eeq

\subsection{Boundary Channel}
\label{3.2}

\begin{figure}
        \centering
        \subfigure[Probe brane boundary channel. \label{fig:boundarya}]{\includegraphics[scale=0.3]{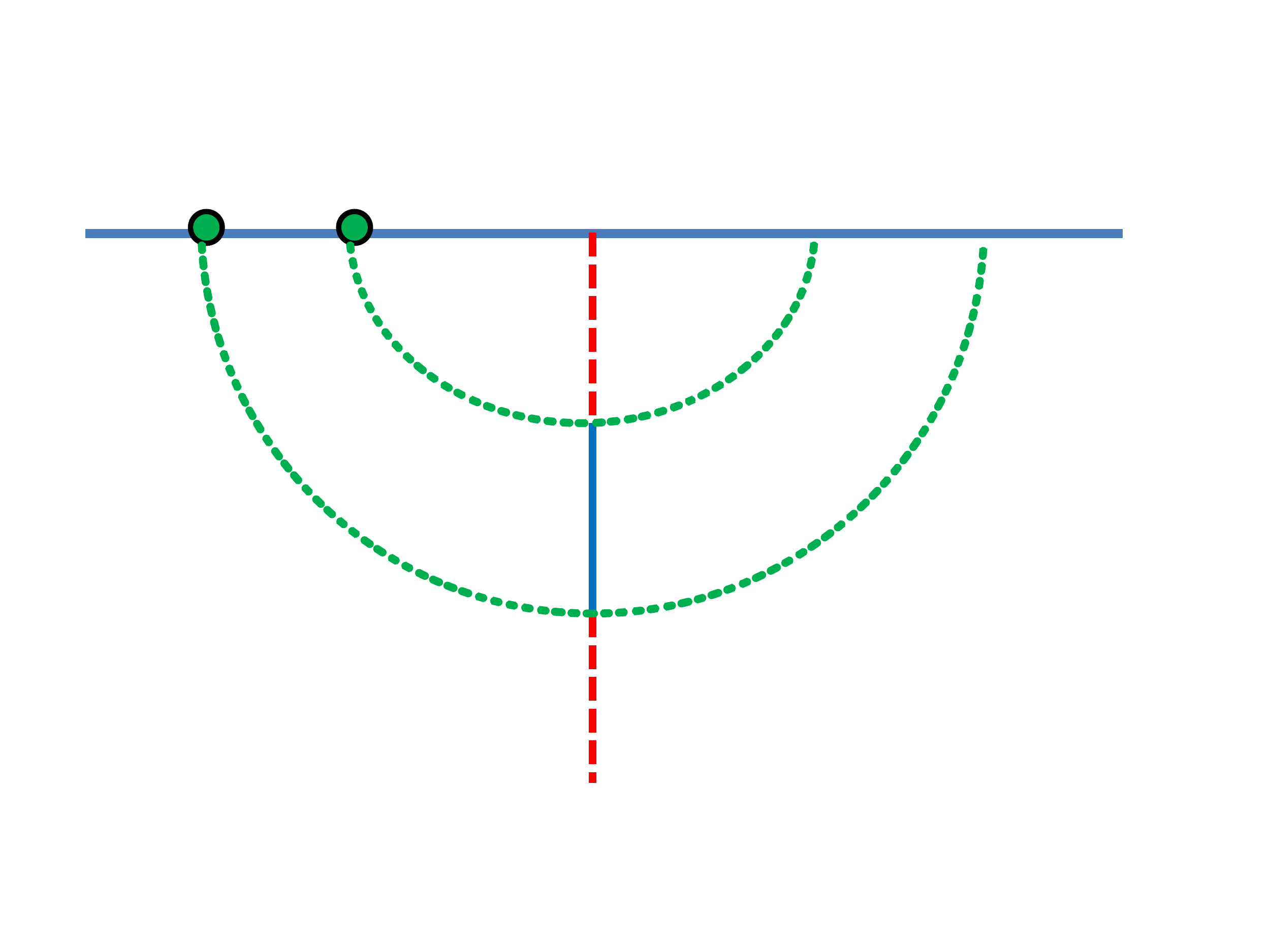}}
        \subfigure[Generic iCFT boundary channel. \label{fig:boundaryb}]{\includegraphics[scale=0.3]{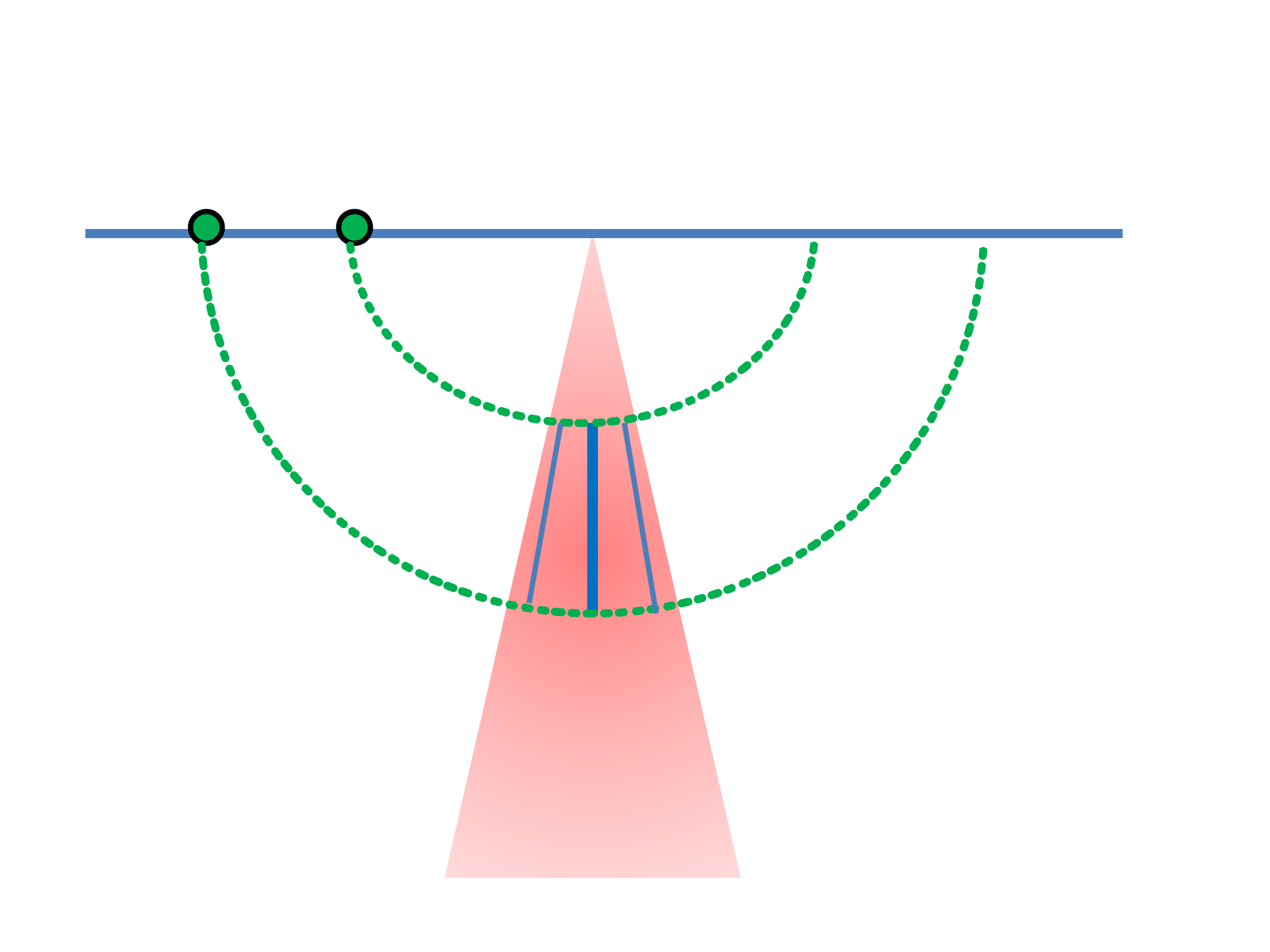}}
		\caption{Boundary Channel Geodesic Witten diagrams.} \label{fig:boundary}
\end{figure}

Despite the simplicity of the Witten diagrams for the full 2-pt function leading to the expression \eqref{twopoint} it
can be helpful to decompose this full answer into a sum over blocks. Here we will use the full form of the 2-pt function to derive an expression in terms of boundary channel blocks. The contribution of a single block will be shown to be given by the diagrams in figure \ref{fig:boundaryb}. In the special case that the holographic bulk dual is simply given by a probe brane our prescription reduces to the Witten diagram of \ref{fig:boundarya} as it appeared previously in \cite{Rastelli:2017ecj}.

To derive the block decomposition, let us start with a representation of the bulk-to-bulk propagator in the full geometry in terms of mode functions. Instead of directly solving \eqref{propeq} for the propagator, we first find a set of appropriate eigenfunctions of the scalar wave operator. Usually one would look for modes of the form
\beq
(\Box_g - M^2(r)) \Phi_{d+1,K}(X) =  E_{K} \Phi_{d+1,K}(X) \,,
\eeq
and then write the Green's function as
\beq G(X_1,X_2)=
\sum_K  \frac{\Phi_{d+1,K}(X_1) \Phi_{d+1,K}(X_2)}{E_K}\,.
\eeq
Here $X$ stands for all the coordinates and $K$ labels the modes. As long as the modes are a complete set, the right hand side automatically gives a delta function when acted upon with the wave operator. In our coordinate system it is easier to follow a slight variation of this strategy. We are looking for modes obeying
\beq
\label{scalarwave}
(\Box_g - M^2(r)) \phi_{d+1,n,k} = \mathrm{e}^{-2A(r)} E_{n,k} \phi_{d+1,n,k}\,.
\eeq
Using the separation of variables ansatz \eqref{sov} together with the form \eqref{eom} of the wave operator these yield
\beq
\label{onslice}
\partial_d^2 \phi_{d,n,k} - m_n^2 \phi_{d,n,k} = E_{n,k} \phi_{d,n,k}
\eeq
instead of our earlier equation \eqref{onslicefree} which we found when looking for solutions of the scalar equation of motion instead of eigenfunctions of the wave operator. Importantly, with our choice to look for solutions of \eqref{scalarwave} the mode equation \eqref{modeeq} remains unchanged and the discrete index $n$ labels its eigenvalues as before. $k$ is labelling the eigenfunctions of the on-slice wave equation \eqref{onslice} for a given $m_n^2$. In the Poincar\'{e} patch slicing of \eqref{ppatch} $k$ is a continuous label. As standard eigenfunctions of the wave equation, $\phi_{d,n,k}$ form a complete set on the AdS$_d$ slice:
\beq
\label{adscomplete}
\int \! \mathrm{d}k \, \phi_{d,n,k}(\vec{x}_1,w_1) \phi_{d,n,k}(\vec{x}_2,w_2) = \frac{\delta(\vec{x}_1-\vec{x}_2)\delta(w_1-w_2)}{\sqrt{-g^0}}
\eeq
where $g^0$ is the determinant of the AdS$_d$ metric on the slice.
In terms of these we can write
\beq
G_{\Delta,d+1}(r_1,\vec{x}_1,w_1,r_2,\vec{x}_2,w_2) = \int \! \mathrm{d}k \sum_n \frac{\phi_{d+1,n,k}(r_1,\vec{x}_1,w_1) \phi_{d+1,n,k}
(r_2,\vec{x}_2,w_2)}{E_{n,k}} \,.
\eeq
Using the defining equation \eqref{scalarwave} as well as completeness \eqref{orthocomplete} and \eqref{adscomplete} of the modes it is straightforward to check that this representation indeed obeys the equation defining the bulk-to-bulk propagator:
\begin{align}
(\Box_g - M^2(r_1) ) & G_{\Delta,d+1}(r_1,\vec{x}_1,w_1,r_2,\vec{x}_2,w_2)  \nonumber \\ &=  ( \Box_g - M^2(r_1) )\int \! \mathrm{d}k \, \sum_n \frac{\phi_{d+1,n,k}(r_1,\vec{x}_1,w_1) \phi_{d+1,n,k}
(r_2,\vec{x}_2,w_2)}{E_{n,k}}   \nonumber \\ &= \mathrm{e}^{-2A(r_1)} \int \! \mathrm{d}k\,  \sum_n \phi_{d+1,n,k}(r_1,\vec{x}_1,w_1) \phi_{d+1,n,k}
(r_2,\vec{x}_2,w_2) \nonumber  \\
&= \nonumber \sum_n \mathrm{e}^{-2A(r_1)}  \psi_n(r_1) \psi_n(r_2) \times \left[ \int \! \mathrm{d}k \, \phi_{d,n,k}(\vec{x}_1,w_1) \phi_{d,n,k}(\vec{x}_2,w_2) \right]
\\
&= \nonumber \frac{ \delta(\vec{x}_1 - \vec{x}_2) \delta(w_1 - w_2)}{\sqrt{-g^0}} \frac{\delta(r_1-r_2)}{\mathrm{e}^{d A}}\\
&= \frac{\delta(r_1-r_2) \delta(\vec{x}_1 - \vec{x}_2) \delta(w_1-w_2)}{\sqrt{-g}}\,.
\end{align}
An important observation is to note that this representation can be used in order to re-express the full $d+1$ dimensional bulk-to-bulk propagator in terms of $d$ dimensional bulk-to-bulk propagators $G^0_{\Delta_n,d}$ of the modes on the slice (that is, these are propagators for a $d$ dimensional scalar on AdS$_d$ with mass squared given by $m_n^2$). The $0$ superscript here is reminding us that, unlike $G_{\Delta,d+1}$, these $d$ dimensional propagators on the slice are calculated using an AdS$_d$ geometry and so are completely known. Given a mode representation of $G^0_{\Delta_n,d}$
\beq
G^0_{\Delta_n,d}(\vec{x}_1,w_1,\vec{x}_2,w_2) = \int \! \mathrm{d}k\, \frac{\phi_{d,n,k}(\vec{x}_1,w_1) \phi_{d,n,k}
(\vec{x}_2,w_2)}{E_{n,k}}
\label{eq331}
\eeq
and the separation of variables ansatz \eqref{sov} we can easily see that the full $d+1$ dimensional propagator can also be represented as\footnote{As a final check of our construction,
note that it is easy to confirm that the propagator in this representation indeed obeys the defining equation \eqref{propeq}:
\begin{align*}
&(\Box_g - M^2) G_{\Delta,d+1} = ({\cal D}^2_{r} +\mathrm{e}^{-2A} \partial_d^2 - M^2) \sum_n \psi_n(r_1) \psi_n(r_2) G^0_{\Delta_n,d}  \\
&=\sum_n \psi_n(r_1) \psi_n(r_2) \left [ G^0_{\Delta_n,d} (- \mathrm{e}^{-2A} m_n^2 + \mathrm{e}^{-2 A} m_n^2)
+ \mathrm{e}^{-2A} \frac{\delta(\vec{x}_1 - \vec{x}_2) \delta(w_1-w_2)}{\sqrt{-g^0}} \right ] \\ \nonumber
&=
\frac{\delta(\vec{x}_1 - \vec{x}_2) \delta(w_1-w_2) \delta(r_1-r_2)}{\sqrt{-g}}
\end{align*}
where we used the completeness of the radial modes \eqref{orthocomplete} as well as
\begin{equation*}
 (\partial_d^2 -m_n^2) \, G^0_{\Delta_n,d} = \frac{\delta(\vec{x}_1 - \vec{x}_2) \delta(w_1-w_2)}{\sqrt{-g^0}}\, .
\end{equation*}
}
\begin{equation}
G_{\Delta,d+1}(r_1,\vec{x}_1,w_1,r_2,\vec{x}_2,w_2) = \sum_n \psi_n(r_1) \psi_n(r_2) G^0_{\Delta_n,d}(\vec{x}_1,w_1,\vec{x}_2,w_2) \,.
\label{propagator}
\end{equation}
Our claim is that this is exactly the decomposition into boundary blocks. This immediately follows from the prescription of \cite{Karch:2017fuh} for identifying the blocks. The bulk-to-bulk propagator is the correlation function of two bulk scalar insertions. We can perform a weighted X-ray transform \eqref{xray} on both insertion points to extract the contribution of a given block:
\beq
f_{\partial,n} \sim {\cal G}_n \equiv \int_{\gamma_1} \! \mathrm{d}r_1 \int_{\gamma_2} \! \mathrm{d}r_2 \, \mathrm{e}^{(d-2)(A(r_1)+A(r_2))}\psi_n(r_1) \psi_n(r_2) G_{\Delta,d+1}(r_1,\vec{x}_1,w_1,r_2,\vec{x}_2,w_2)\, .
\eeq
Furthermore the orthogonality of the mode-functions \eqref{orthocomplete} yields
\beq
{\cal G}_n =   G^0_{\Delta_n,d}(\vec{x}_1,w_1,\vec{x}_2,w_2) \, .
\eeq
That is, the conformal boundary block is supposed to be equal to the AdS$_d$ bulk-to-bulk propagator up to some normalization constant which we will determine shortly. The explicit form of the AdS$_d$ bulk-to-bulk propagator in the conventions of \cite{Ammon:2015wua} is
\beq
G^0_{\Delta_n,d} = \frac{C_{\Delta_n,d}}{2^{\Delta_n} (2 \Delta_n-(d-1))} \xi^{\Delta_n} \,
{}_2 F_1 \left(\frac{\Delta_n}{2},\frac{\Delta_n+1}{2},\Delta_n - \frac{d-1}{2}+1,\xi^2\right)
\eeq
with
\beq
\label{normfactors}
C_{\Delta_n,d} = \frac{\Gamma(\Delta_n)}{\pi^{(d-1)/2} \Gamma(\Delta_n - \frac{d-1}{2})}
\eeq
and $\xi$ the chordal distance, which in the coordinates of \eqref{ppatch} becomes
\beq
\xi= \frac{2 w_1 w_2}{w_1^2 + w_2^2 + (\vec{x}_1 - \vec{x}_2)^2}
\eeq
or in other words
\beq
\xi=\frac{2}{\eta+2} = \frac{-4/\eta}{-4/\eta -2} \,.
\eeq
Using a quadratic hypergeometric identity\footnote{ $${}_2 F_1(a,b,2b,z) = \left(1-\frac{z}{2}\right)^{-a}
 {}_2 F_1(a/2,a/2+1/2,b+1/2,[z/(2-z)]^2)\,.$$.} we can also write this as
\beq
G^0_{\Delta_n,d} = \frac{C_{\Delta_n,d}}{ 4^{\Delta_n} (2 \Delta_n-(d-1))} \, \left ( \frac{\eta}{4} \right )^{- \Delta_n} \,
{}_2 F_1 \left(\Delta_n,\Delta_n-\frac{d}{2}+1,2\Delta_n - d +2, - \frac{4}{\eta}\right) \,.
\eeq
Comparing with the expression for $f_{\partial}$ \eqref{boundaryblock} we see that they indeed have exactly the same functional form. This can also easily be verified by the Casimir method \cite{Rastelli:2017ecj}; the form of the block is entirely fixed by conformal invariance.
To confirm that the ${\cal G}_n$ we obtained via the weighted X-ray transform indeed contribute to the CFT correlation functions exactly like a boundary block we need to plug in our representation of the propagator \eqref{propagator}
into the formula for the full 2-pt function \eqref{twopoint}. To do so we need the asymptotic behavior of the mode functions $\psi_n(r)$. According to the analysis in \cite{Aharony:2003qf} at large $r$ one finds
\beq
\psi_n = C_n (\mathrm{e}^r)^{-\Delta} + {\cal O}(\mathrm{e}^{-(\Delta+2)r})
\eeq
that is the fall-off is universally determined by the mass $M_0^2$ of the bulk scalar irrespective of $n$. The numerical factors $C_n$ are related to the BOPE coefficients as we will also make explicit below.
Correspondingly we find for the 2-pt function
\beq
\label{bounddeco}
\langle O(\vec{x}_1,w_1) O(\vec{x}_2,w_2) \rangle = \frac{C (2\Delta-d)^2}{(2 w_1)^{\Delta} (2 w_2)^{\Delta}} \, \sum_n C_n^2 \,  {\cal G}_n \,.
\eeq
Collecting all the numerical factors and comparing with the desired boundary channel decomposition \eqref{boundaryf} one finds
\beq
\label{coeffs}
(c_n^O)^2 = (C_n)^2 C (2 \Delta -d)^2 \frac{ C_{\Delta_n,d}}{2 \Delta_n - (d-1)} \frac{1}{4^{\Delta_n}}. \eeq
Recall that $C$ was the prefactor of the action, $C_n$ the coefficient governing and fall-off of the mode functions, $C_{\Delta_n,d}$ the standard normalization factors \eqref{normfactors} appearing in the bulk-to-bulk propagator and $c_n^O$ the BOPE coefficient. The factor of $(2 \Delta -d)^2$ is inherited from the relation between correlation function and bulk-to-bulk propagator \eqref{twopoint}.

The prescription of \eqref{bounddeco} with \eqref{coeffs} provides a full non-perturbative decomposition of the 2-pt function into conformal blocks valid in any bCFT with holographic dual. It constitutes the main result of our work. If we want to describe the bCFT in terms of Witten diagrams in which the effect of the defect itself is treated order by order in a diagrammatic expansion,  we need to look for a situation where the effects of the defect/interface can be taken into account perturbatively so that a calculation of the wavefunctions order by order in a diagrammatic expansion is meaningful to begin with. That is, we are interested in a situation in which the metric and background fields obey
\beq \label{small} g=g_{\mathrm{AdS}_{d+1}} + \varepsilon^2 \, \delta g, \quad \quad  X= \varepsilon \, \delta X(r) \eeq
with $\varepsilon$ a small parameter\footnote{As long as the stress tensor is quadratic in $X$ the correction to the metric is of order $\varepsilon^2$. This is generically the case but the analysis in this section can also easily be adapted to the case where this assumption fails.}.
The mode functions encode the bulk geometry via the analog Schr\"odinger equation with potential \eqref{analog} as emphasized in \cite{Karch:2017fuh}. As long as the background has the form of \eqref{small}, the potential fixing the Hamiltonian of the analog Schr\"odinger system takes the form
\beq
V=V_0 + \varepsilon \, \delta V
\label{potential}
\eeq
where $V_0$ is the potential associated with an AdS$_{d+1}$ geometry.
Correspondingly we can expand the energy and the eigenfunctions as
\beq
m_n^2= (m_n^0)^2+\varepsilon \, \delta m_n^2 \quad \mbox{and} \quad
\psi_n = \psi_n^0 + \varepsilon \, \delta \psi_n \,.
\eeq
We will study the eigenfunctions of $V_0$ in all detail in the next section, for now it suffices to say that, using our result from \eqref{propagator} the propagator,
\beq
\label{adspropagator}
G^0_{\Delta,d+1}(r_1,\vec{x}_1,w_1,r_2,\vec{x}_2,w_2) = \sum_n \psi^0_n(r_1) \psi^0_n(r_2) G^0_{\Delta_n,d}(\vec{x}_1,w_1,\vec{x}_2,w_2).
\eeq
is just a non-standard representation of the usual AdS$_{d+1}$ bulk-to-bulk propagator.

Now let us turn to the leading correction to the 2-pt function at small $\varepsilon$. Since $\psi_n=\psi_n^0 + \varepsilon \, \delta \psi_n$ we also have $C_n=C_n^0 + \varepsilon \, \delta C_n$. According to standard quantum mechanical perturbation theory we can write
\begin{equation}
\frac{\delta m_n^2}{2}= \int \! \mathrm{d}r \, \mathrm{e}^{(d-2)A} \psi_n^0(r) \delta V (r) \psi_n^0 (r)
\end{equation}
and
\beq
\delta \psi_n = \sum_{m \neq n} \gamma_{mn} \psi_m^0 \quad \mbox{and} \quad \delta C_n = \sum_{m \neq n} \gamma_{mn} C_m^0
\eeq
with
\beq
\gamma_{mn} = 2 \int \! \mathrm{d}r \, \frac{\mathrm{e}^{(d-2) A(r)} \psi_m^0(r) \delta V(r) \psi_n^0(r) }{(m_n^0)^2-(m_m^0)^2}\,.
\eeq
The leading correction to the 2-pt function hence becomes
\begin{equation}
\begin{split}
\label{bounddecosmall}
&\delta \langle O(\vec{x}_1,w_1) O(\vec{x}_2,w_2) \rangle \\
 & \quad \quad \quad \quad =\varepsilon \frac{C (2\Delta-d)^2}{(2 w_1)^{\Delta} (2 w_2)^{\Delta}}
\left(2  \sum_{n,m \neq n} \gamma_{mn} C_m^0 C_n^0 \,  {\cal G}_n +\sum_n \delta m_n^2 (C_n^0)^2 \tilde{\mathcal{G}}_n\right)
\end{split}
\end{equation}
with
\begin{equation}
\tilde{\mathcal{G}}_n=\frac{G_{\Delta_n^0+\varepsilon \delta \Delta_n,d}^0-G_{\Delta_n^0,d}^0}{\varepsilon \, \delta m_n^2} \,.
\end{equation}
Eq. \eqref{bounddecosmall} can be read as our result for the geodesic Witten diagram prescription for calculating the contribution of a given boundary block to the scalar 2-pt function. Symmetry enhanced geodesics are used to identify points on the AdS$_d$ slices associated to a given boundary point. The first term corresponds to a diagram that represents the bulk-to-bulk propagator on the slice weighted by $\sum_{m\neq n}\gamma_{mn} C_m^0 C_n^0$. The second term is proportional to the correction of the propagator on the slice due to the shifted mass. Diagrammatically this could be represented by a mass insertion on the slice connected with two propagators, even though it is not clear this representation would be very illuminating. Note that the integrand of  $\delta m_n^2$ and $\gamma_{mn}$ is only non-vanishing in the region where the geometry differs from AdS$_{d+1}$ as depicted in figure \ref{fig:boundaryb}.

The case of probe brane considered in \cite{Rastelli:2017ecj} and depicted in figure \ref{fig:boundarya} looks similar but different in detail. In \cite{Rastelli:2017ecj} the on-slice propagator is weighted by two bulk-to-boundary propagators as well as two interaction vertices. We will now show that this result in fact also naturally arises from our more general prescription. What is new in the probe brane case is that we have in the holographic theory matter fields which are completely localized on the brane. They are dual to defect localized operators in the CFT which do not arise from restriction of an ambient space operator to the defect but instead arise from matter that only lives on the defect. In terms of our bulk prescription such brane localized matter fields correspond to special modes $\psi_M$ of dimension $\Delta_M$ which, before accounting for the interactions, have no support near the boundary, meaning $C^0_M=0$. As such, they make no contribution to the zeroth order 2-pt function. Since $C_M^0$ vanishes, we also have $\delta C_n = \gamma_{nM} C_M^0=0$. The leading contribution of these extra field comes from the correction to $\psi_M$ (and hence $C_M$):
\beq
\psi_M = \psi_M^0 + \sum_n \varepsilon \, \gamma_{nM} \psi_n^0 = \psi_M^0 + \varepsilon \, \delta \psi_M.
\eeq
$\psi_M^0$ has no support near $r \rightarrow \infty$ and so does not contributed to correlation functions. The leading contribution of this mode to the 2-pt function hence is
\beq
\label{probe}
\delta \langle O(\vec{x}_1,w_1) O(\vec{x}_2,w_2) \rangle = \varepsilon^2 \frac{C (2\Delta-d)^2}{(2 w_1)^{\Delta} (2 w_2)^{\Delta}} \, (\delta C_M)^2\,  {\cal G}_M \,.
\eeq
This result appears already much closer to the answer quoted in \cite{Rastelli:2017ecj} in that it is order $\varepsilon^2$, that is it involves two coupling insertions, and is proportional to the propagator of a brane localized field which did not contribute to the zeroth order answer. In \cite{Rastelli:2017ecj} this propagator was decorated with two bulk-to-boundary propagators of the bulk field dual to the operator of dimension $\Delta$. These bulk-to-boundary propagators connected the boundary operator insertion point with coordinates $(\vec{x},w)$ to the bulk point $(r=0,\vec{x},w)$ to which it is tied via a symmetry enhanced geodesic. For this special case the bulk-to-boundary propagators simply amount to inserting\footnote{The bulk to boundary propagator on AdS$_{d+1}$ in the standard Poincar\'{e} coordinates with $\mathrm{d}s^2=W^{-2}(\eta_{MN} \mathrm{d}X^M \mathrm{d}X^N)$ is proportional to $W^{\Delta}/(W^2 + (\vec{X}_1-\vec{X}_2)^2)^{\Delta}$. We can change variables to $r,w,\vec{x}$ by setting
\beq \cosh(r) w^{-1} = W^{-1}, \quad w \tanh(r) = X_{d-1}, \quad \vec{x} = \vec{X}\,. \eeq
With this the boundary point $(w,\vec{x})$ lives at $(X_{d-1}=w,\vec{X}=\vec{x})$ where as the bulk point $(r=0,w,\vec{x})$ at $(W=w,X_{d-1}=0,\vec{x})$. The bulk-to-boundary propagator connecting the two indeed just gives a factor of $(2w)^{-\Delta}$.}
the factors of $(2w_1)^{\Delta}$ and $(2 w_2)^{\Delta}$ which arise automatically in our answer \eqref{probe}. So our answer is indeed completely equivalent to the one found in \cite{Rastelli:2017ecj}.

\subsubsection{An explicit example: The Holographic No-braner}
\label{3.2.1}

The simplest theory to test our decomposition \eqref{bounddeco} on is the holographic no-braner. That is, as described in subsection \ref{examples}, we take our field theory to be a field theory without boundary with the plane at $w=0$ treated as a defect. In this case the BOPE simply becomes a Taylor series. It was already found in \cite{Aharony:2003qf} that this seemingly trivial example appears actually quite non-trivial from the point of view of the BOPE. Here we will see that also from the point of view of the conformal block expansion we require some seemingly miraculous cancellations.

Let us start from the field theory point of view. Consider an operator $O$ of dimension $\Delta$ in a $d$ dimensional CFT without any brane, boundary or defect. In this case, the full $SO(d,2)$ conformal group demands that
\beq
\langle O (\vec{x}_1,w_1) O(\vec{x}_2,w_2) \rangle = \frac{ {\cal N} }{[(\vec{x}_1 - \vec{x}_2)^2 + (w_1-w_2)^2]^{\Delta}} \,.
\eeq
Here ${\cal N}$ is an overall normalization constant which could be chosen to be 1.
Comparing this with the form \eqref{form} of the 2-pt function in a bCFT we see that this corresponds to
\beq
\label{fnobrane}
 f_{\text{no-brane}}(\eta)= {\cal N} \left ( \frac{\eta}{4} \right )^{-\Delta}. \eeq
This point is obvious from the ambient channel decomposition of the 2-pt function in \eqref{bulkf}. In the ``no-braner" theory only the identity operator has a non-trivial 1-pt function and so this is the only block that contributes. From the boundary channel point of view, we however have an infinite tower of boundary operators with dimension $\Delta+n$ for $n=0,1,2,\ldots$ contributing to the 2-pt function. The fact that $O$ doesn't have a vev means there is no contribution from the identity operator in the boundary channel. The coefficients $c_n^O$ in the decomposition \eqref{boundaryf} have to conspire in such a way that the contribution of all the blocks sums up to the simple expression in \eqref{fnobrane}. In the special case of a free no-brane theory we saw that only two boundary operators contributed, but for a general bCFT (in particular one with a holographic dual) we genuinely need an infinite tower to sum up into the simple power law we are looking for.

This simple example can then be seen as one explicit solution to the conformal boundary bootstrap where an infinite number of boundary blocks reproduces the ambient channel decomposition in which only the identity contributes:
\beq
{\cal N} \left ( \frac{\eta}{4} \right )^{-\Delta} = \sum_{n=0}^{\infty} (c_n^O)^2 f_{\partial,n}
\label{357}
\eeq
with $f_{\partial,n}$ given by \eqref{boundaryblock} with dimension $\Delta_n=\Delta+n$. This requires a rather involved identity of hypergeometric functions and provides a non-trivial check for the holographic calculation of the coefficients in \eqref{coeffs}.
Below we will prove the identity for the special case that $d=\Delta=4$, but we expect \eqref{357} to hold for general $d$ and $\Delta$\footnote{The identity \eqref{357} is a special case of eq. (A.7b) in \cite{Hogervorst:2017kbj}.
If one inserts $h=d/2$, $\ell_1=\ell_2=-\Delta$ and $\rho =\eta /(\eta+4)$ into eq. (A.7b) in  \cite{Hogervorst:2017kbj}, one will obtain \eqref{357}. 
Closely related identities have also been used in \cite{Herzog:2017xha}.
We would like to thank Christopher Herzog for pointing out \cite{Hogervorst:2017kbj,Herzog:2017xha} to us.}. 

The dynamics of a holographic bCFT is encoded in the bulk geometry via the warpfactor $\mathrm{e}^A$ as well as the non-trivial background fields $X(r)$. As emphasized above, the only place this data enters into the BOPE coefficients in \eqref{coeffs} is via the $C_n$, that is via the asymptotic fall-off of the mode functions. For the no-brane theory, the geometry is AdS$_{d+1}$, which means $\mathrm{e}^A=\cosh(r)$, and all other background fields are turned off. In this case the mode equation \eqref{modeeq} can be solved analytically \cite{Karch:2000ct,Porrati:2001gx}. For $d=\Delta=4$ one finds
\beq
\psi_n \sim (\cosh^{-4}r) \, {}_2 F_1 \left(\frac{5}{2} + \frac{n}{2},-\frac{n}{2},3,\cosh^{-2} r\right)\,.
\eeq
The overall prefactor can be determined by requiring the orthogonality condition \eqref{orthocomplete}. The asymptotic fall-off can now simply be read off and one finds\footnote{Ref. \cite{Aharony:2003qf} chose to work with un-normalized wavefunctions and an $n$-independent fall-off $C_n$. In this case the non-trivial coefficient was obtained from the norm $C_n^{-2} \sim \int \! \mathrm{d}r \, \mathrm{e}^{2A} \psi_n^2$.}
\cite{Aharony:2003qf}
\beq
(C_n)^2 = 2 \frac{(2n+5) (n+4)!}{n!} \,.
\eeq
Collecting all the prefactors in \eqref{coeffs} and grouping all $n$-independent coefficients into an overall constant $\tilde{\alpha}$ we find
\beq
(c_n^O)^2 = \tilde{\alpha} \frac{(2n+5) (n+4)!}{n!} \frac{\Gamma(n+4)}{\Gamma(n+\frac{5}{2})} \frac{1}{(2n+5) 4^n}\,.
\eeq
We are hence tasked to calculate the sum (using $z=-4/\eta$ for simplicity)
\beq
f\left(-\frac{4}{z}\right) =  \tilde{\alpha} \, \sum_{n=0}^{\infty}  \frac{(2n+5) (n+4)!}{n!} \frac{\Gamma(n+4)}{\Gamma(n+\frac{5}{2})} \frac{ (-z)^{n+4} }{(2n+5) 4^n}  \,  {}_2 F_1(n+4,n+3,2n+6, z )\,.
\eeq
The claim is that this has to reproduce the simple power law from \eqref{fnobrane}.
We are not aware of any known sum formula for hypergeometric functions of this kind, but it is actually quite straight forward to show that this is indeed the case. We start by using an integral definition for the hypergeometric function\footnote{The general expression is $${}_2 F_1(a,b,c,z) = \frac{\Gamma(c)}{\Gamma(b) \Gamma(c-b)} \, \int_0^1 \mathrm{d}t \, t^{b-1} (1-t)^{c-b-1} (1-zt)^{-a}\,.$$.}
to write
\beq
(-z)^{n+4} \, {}_2 F_1(n+4,n+3,2n+6, z ) = \frac{\Gamma(2n+6)}{\Gamma(n+3)^2} \,
z^2 \int_0^1 \mathrm{d}t \,  \left ( z \frac{ t (t-1)}{1-zt} \right )^{n+2} \frac{1}{(1-zt)^2}\,.
\eeq
This allows us to write (using $z=-4/\eta$)
\beq
f(z) = \tilde{\alpha} z^2 \int_0^1 \! \mathrm{d}t \, \frac{1}{(1-zt)^2} \, \left [  \sum_n \beta_n \left ( z \frac{ t (t-1)}{1-zt} \right )^{n+2}    \right ]
\eeq
where the coefficients $\beta_n$ are given by\footnote{We can simplify the product of Gamma functions by using the
Legendre Duplication formula $$\Gamma(2z) = \frac{2^{2z-1}}{\pi^{1/2}} \Gamma(z) \Gamma(z+1/2)$$
on $\Gamma(2n+6)$.}
\beq
\beta_n =(2n+5) (n+4) (n+3)^2 (n+2) (n+1)\,.
\eeq
Now the sum over $n$ is straightforward to do and so is the subsequent integral over $t$. We find, as hoped for,
\beq
f(\eta) = 12 \tilde{\alpha} \left ( \frac{4}{\eta}  \right )^4
\eeq
in perfect agreement with \eqref{fnobrane}.

\subsection{Ambient Channel}
\label{3.3}

\begin{figure}
        \centering
       {\includegraphics[scale=0.3]{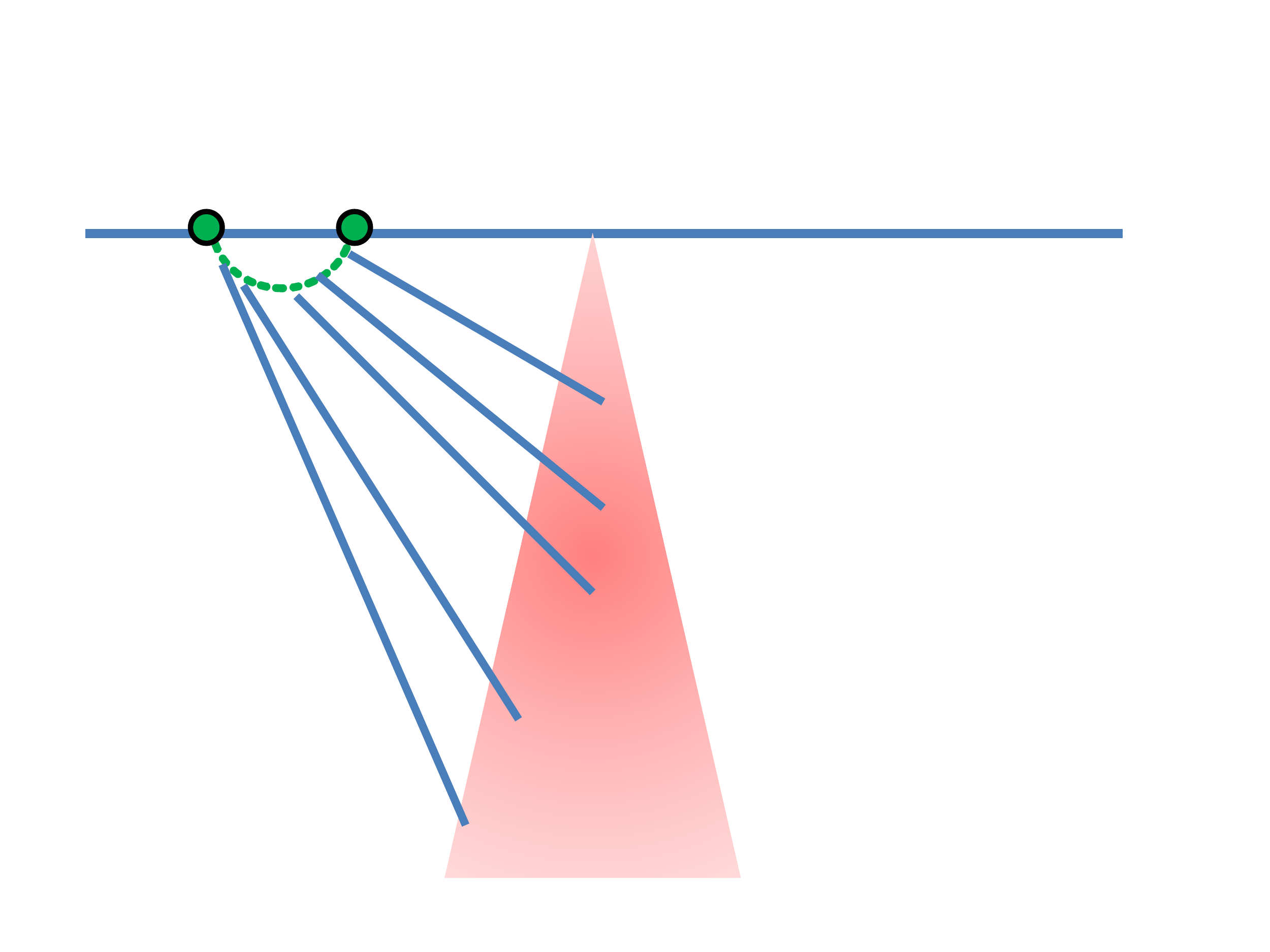}}
		\caption{Generic contribution to the full 2-pt function.} \label{fig:ambientjanusfull}
\end{figure}

The boundary channel expansion relied heavily on a new feature of bCFTs: the BOPE. In contrast, the ambient channel expansion uses the standard OPE for operators. The defect only makes its presence known via allowing non-vanishing 1-pt functions for scalar operators. Correspondingly the structure of the ambient channel expansion closely follows the pattern in the theory without defects \cite{Hijano:2015zsa}. First note that we can always think of the non-trivial metric \eqref{metric} in the framework of being AdS$_{d+1}$ plus deformations, as formally displayed in \eqref{small} with $\varepsilon$ not necessarily small. The simple Witten diagram for the full 2-pt function of figure \ref{fig:full} can be thought of as an infinite sum of Witten diagrams
in AdS$_{d+1}$ with $\delta g$ and $\delta X$ insertions. A generic contribution is displayed in figure \ref{fig:ambientjanusfull}.

\begin{figure}
        \centering
        \subfigure[Probe brane ambient channel. \label{fig:ambienta}]{\includegraphics[scale=0.3]{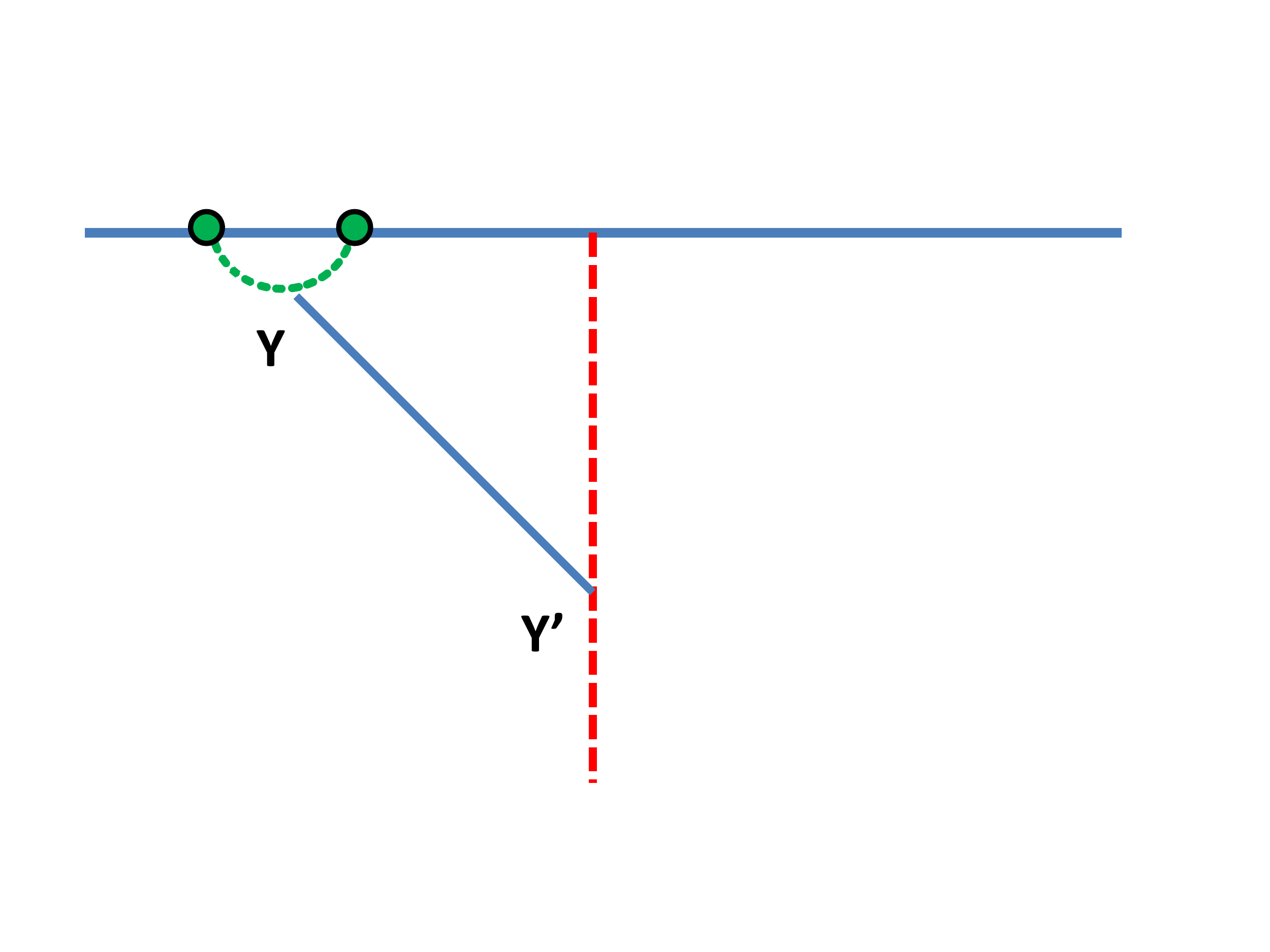}}
        \subfigure[Generic iCFT ambient channel. \label{fig:ambientb}]{\includegraphics[scale=0.3]{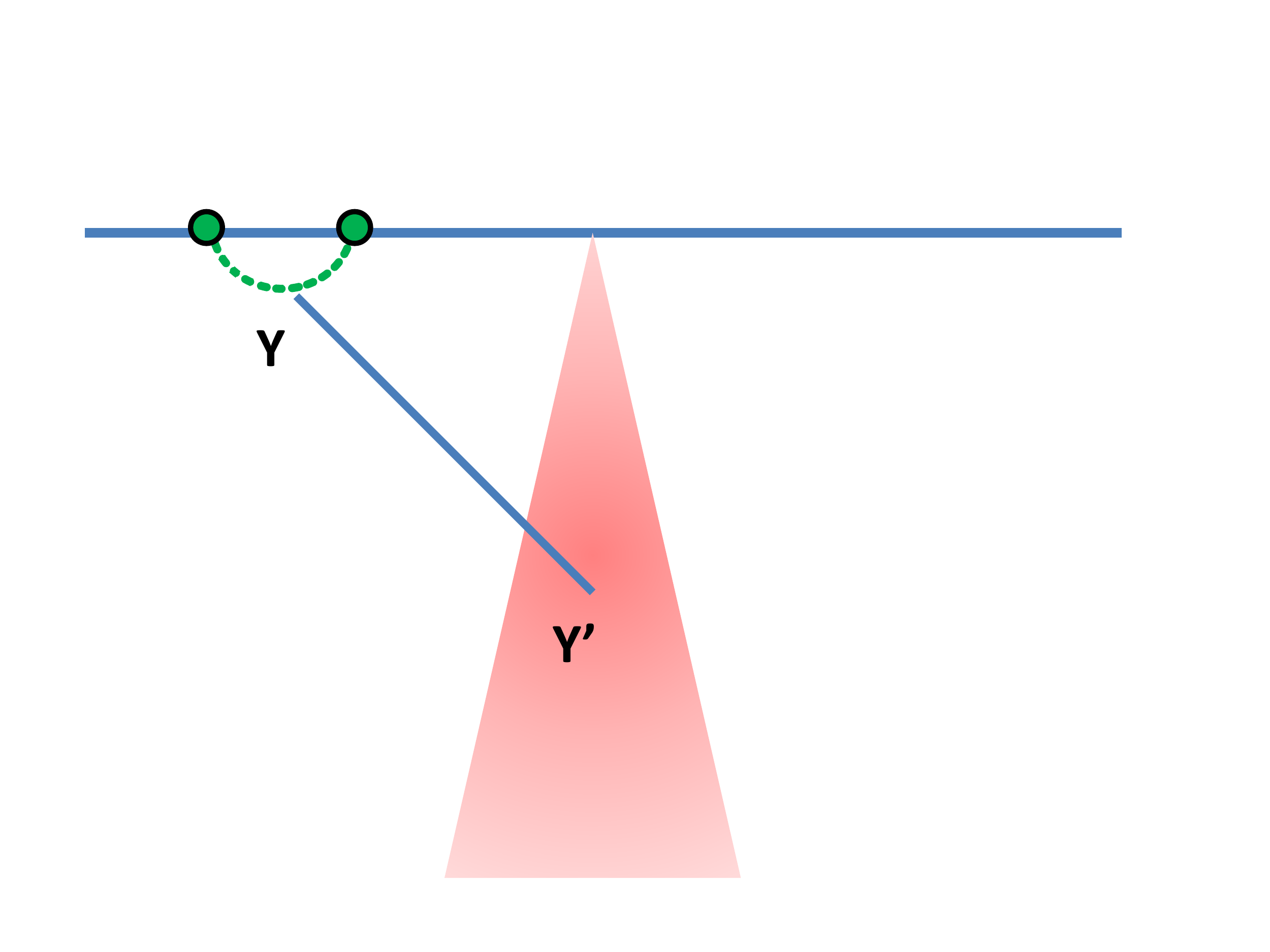}}
		\caption{Ambient Channel Geodesic Witten diagrams.} \label{fig:ambient}
\end{figure}

For any expansion in terms of Witten diagrams to make sense, we need to focus on the case of small $\varepsilon$.
In this case we consider diagrams in pure AdS$_{d+1}$ and treat the deviations of the warpfactor from the $\mathrm{e}^A=\cosh(r)$ form as well as all matter fields with non-trivial profile $X(r)$ as extra sources.
We can derive the ambient channel expansion from the identity
\begin{equation}
G_{\Delta ,d+1} (X,Z)=\int \! \mathrm{d}^{d+1} Y' \sqrt{-g} \, G_{\Delta ,d+1} (X,Y') [\Box_g-M^2] G_{\Delta ,d+1} (Y',Z)\,.
\end{equation}
Here $X$ and $Z$ stand for bulk points $(r_1,\vec{x}_1,w_1)$ and $(r_2, \vec{x}_2,w_2)$ respectively. $Y'$ is a bulk point and integration region for $Y'$ is the whole of AdS$_{d+1}$ spacetime. $\Box_g$ denotes the Laplacian at $Y'$.
Expanding the identity around the background and taking the $r_{1,2}\to \infty$ limit, we obtain
\begin{equation}
\delta \langle O(\vec{x}_1,w_1) O(\vec{x}_2,w_2) \rangle  =
2 C \varepsilon \int \!  \mathrm{d}^{d+1}Y' \sqrt{-g} \,
K^0_{\Delta,d+1}(Y',\vec{x}_1,w_1)  K^0_{\Delta,d+1}(Y',\vec{x}_2,w_2) \, \mathrm{e}^{-2A} \delta V(r')
\label{ambient2}
\end{equation}
where we used $\delta [\Box_g-M^2]=-2\varepsilon \, \mathrm{e}^{-2A}\delta V (r')$.
Here, as before, the 0 superscript indicates that these quantities take their un-deformed AdS$_{d+1}$ values.
The two bulk-to-boundary propagators can be decomposed as \cite{Hijano:2015zsa}
\begin{equation}
\begin{split}
&K^0_{\Delta,d+1}(\vec{x}_1,w_1,Y')  K^0_{\Delta,d+1}(Y',\vec{x}_2,w_2)  \\
&=\sum_N b_N \int _{\tilde{\gamma}_0} \! \mathrm{d} \lambda  \,
K^0_{\Delta,d+1}(\vec{x}_1,w_1,Y(\lambda))  K^0_{\Delta,d+1}(Y(\lambda ),\vec{x}_2,w_2)
G_{\Delta_N,d+1}^0(Y(\lambda),Y')
\end{split}
\end{equation}
with
\[
b_N=
\frac{2\Gamma (\Delta_N)}{\Gamma (\Delta_N/2)^2}
\frac{(-1)^N}{N!} \frac{[(\Delta)_N]^2}{(2\Delta +N-d/2)_N} \qquad \mbox{and} \qquad \Delta_N=2\Delta +2N \,.
\]
Here $(x)_n=\Gamma (x+n)/\Gamma (x)$ represents the Pochhammer symbol.
The geodesic $\tilde{\gamma}_0$ parametrized by $\lambda$ is the usual AdS$_{d+1}$ geodesic connecting the boundary points $(\vec{x}_1,w_1)$ and $(\vec{x}_2,w_2)$.
Plugging this decomposition into \eqref{ambient2}, we obtain conformal block expansions of the ambient channel,
\begin{equation}
\begin{split}
\delta \langle & O(\vec{x}_1,w_1) O(\vec{x}_2,w_2) \rangle    \\
& =2 C \varepsilon \sum_N b_N \int _{\tilde{\gamma}_0} \! \mathrm{d} \lambda \int \!  \mathrm{d}^{d+1}Y' \, \sqrt{-g} \, 
K^0_{\Delta,d+1}(Y(\lambda),\vec{x}_1,w_1)  K^0_{\Delta,d+1}(Y(\lambda ),\vec{x}_2,w_2)  \\
& \qquad \qquad \qquad \qquad \qquad \qquad \qquad \times 
G_{\Delta_N,d+1}^0(Y(\lambda),Y') \, \mathrm{e}^{-2A} \delta V(r')\,.
\label{ambient}
\end{split}
\end{equation}
This shows that there is basically no difference between
a probe brane setup (in which case the sources are localized at a single locus $r=r_*$ as depicted in figure \ref{fig:ambienta}) and the generic holographic bCFT geometry of \eqref{metric} (in which case the sources have support over some region in $r$ as depicted in figure \ref{fig:ambientb}).
Note that this integral only involves the uncorrected AdS$_{d+1}$ propagators.

The proof in \cite{Rastelli:2017ecj} that this is indeed an eigenfunction of the conformal Casimir with the right boundary conditions and hence corresponds to a contribution of a single ambient channel block only relied on the properties of the geodesic $\tilde{\gamma}_0$ and the bulk-to-bulk propagator $G^0_{\Delta,d+1}$. It applies immediately to our case as well.

The one interesting upshot of this analysis is that to leading order the only aspect of the bulk geometry that affects the 2-pt function of two scalar operator $O$ are the bulk scalars $X(r)$ with non-trivial profile. This is due to the fact that only scalar blocks contribute in the 2-pt function of ambient space scalars. This follows immediately from angular momentum conservation.

\subsection{Equivalence of two different decompositions}
\label{3.4}

In subsections \ref{3.2} and \ref{3.3} we decomposed 2-pt functions in two different ways; the boundary channel \eqref{bounddecosmall} and the ambient channel \eqref{ambient}.
These decompositions should be the same.
In this subsection we give an explicit proof of this equivalence.
At leading order we confirmed this in subsection \ref{3.2.1} when discussing the no-brane case.

Our mode decomposition of the AdS$_{d+1}$ propagator \eqref{adspropagator} implies a similar representation for the bulk-to-boundary propagator via the limiting procedure of \eqref{bulktobound}. Using this representation for the bulk-to-boundary propagators in our ambient channel result \eqref{ambient2}, we get
\begin{equation}
\begin{split}
& \delta \langle O(\vec{x}_1,w_1) O(\vec{x}_2,w_2) \rangle  =   \frac{2 C \varepsilon (2\Delta -d)^2}{(2w_1)^\Delta (2w_2)^\Delta}\int  \! \mathrm{d}r' \, \mathrm{e}^{(d-2)A} \int \! \mathrm{d}\vec{x}' \mathrm{d}w' \, \sqrt{-g^0} \\
& \times \sum_{n,m}C_n^0C_m^0 \psi _n^0(r') \psi_m^0 (r') \delta V(r') G_{\Delta_n^0,d}^0 (\vec{x}_1,w_1,\vec{x}',w')G_{\Delta_m^0,d}^0 (\vec{x}',w',\vec{x}_2,w_2)
\end{split}
\end{equation}
Furthermore, using the usual expression for first order perturbation theory for the analog Schr\"odinger equation

\begin{equation}
2\int \! \mathrm{d} r' \, \mathrm{e}^{(d-2)A}\psi _n^0(r') \psi_m^0 (r') \delta V(r')=
\begin{cases}
 \delta m_n^2  &\mbox{for } n= m \\   
((m_n^0)^2-(m_m^0)^2)\gamma _{mn}  & \mbox{for } n\neq m
\end{cases}
\end{equation}
and a slightly reorganized equation of motion for the propagator
\beq (m_n^0)^2 G_{\Delta_n^0,d}^0 (\vec{x}_1,w_1,\vec{x}',w') =\partial_d^2 G_{\Delta_n^0,d}^0 (\vec{x}_1,w_1,\vec{x}',w') -\frac{1}{\sqrt{-g^0}}\delta (\vec{x}_1-\vec{x}')\delta (w_1-w') \eeq
we obtain
\begin{align}
\delta \langle O(\vec{x}_1,w_1) & O(\vec{x}_2,w_2) \rangle  =\varepsilon \frac{C (2\Delta-d)^2}{(2 w_1)^{\Delta} (2 w_2)^{\Delta}} \left( 2 \sum_{n,m \neq n} \gamma_{mn} C_n^0 C_m^0 \,  {\cal G}_n \right.  \\
&\left. +\sum_n (C_n^0)^2\delta m_n^2 \int \! \mathrm{d}\vec{x}' \mathrm{d}w' \, \sqrt{-g^0}\,  G_{\Delta_n^0,d}^0 (\vec{x}_1,w_1,\vec{x}',w')G_{\Delta_n^0,d}^0 (\vec{x}',w',\vec{x}_2,w_2)  \right)
\,. \notag
\end{align}
The first term is perfect agreement with that of the leading correction of boundary channel \eqref{bounddecosmall}.
To show the equivalence of the second term, some further computations are required.
Using \eqref{eq331} and the completeness relation of $\phi_{d,n,k}$, we obtain\footnote{In terms of diagrams what we are saying is that the change of the AdS$_d$ propagator of a field with shifted mass $(m_n^0)^2 + \varepsilon \, \delta m_n^2$ can be obtained from a Witten diagram with an interaction vertex $\delta m_n^2$, integrated over all of AdS$_d$, connecting two propagators associated to mass $(m_n^0)^2$.}
\begin{align}
&\varepsilon \, \delta m_n^2 \int \! \mathrm{d}\vec{x}' \mathrm{d}w' \, \sqrt{-g^0}\,  G_{\Delta_n^0,d}^0 (\vec{x}_1,w_1,\vec{x}',w')G_{\Delta_n^0,d}^0 (\vec{x}',w',\vec{x}_2,w_2)  \notag \\
& =\varepsilon \, \delta m_n^2 \int \! \mathrm{d}^d y' \, \sqrt{-g^0}
\int \! \mathrm{d} k\,\frac{\phi_{d,n,k}(y_1) \phi_{d,n,k}(y') }{E_{n,k}}
\int \! \mathrm{d} \ell \, \frac{\phi_{d,n,\ell}(y') \phi_{d,n,\ell }(y_2) }{E_{n,\ell}} \notag \\
&= \int \mathrm{d}k \,  \frac{\phi_{d,n,k}(y_1) \phi_{d,n,k}(y_2)}{E_{n,k}}\cdot \frac{\varepsilon \, \delta m_n^2}{E_{n,k}} \notag \\
&= \int \mathrm{d}k \,  \frac{\phi_{d,n,k}(y_1) \phi_{d,n,k}(y_2)}{E_{n,k}-\varepsilon \, \delta m_n^2} -\int \mathrm{d}k \,  \frac{\phi_{d,n,k}(y_1) \phi_{d,n,k}(y_2)}{E_{n,k}} \notag \\
&=G_{\Delta_n^0+\varepsilon \delta \Delta_n,d}^0(\vec{x}_1,w_1,\vec{x}_2,w_2) -G_{\Delta_n^0,d}^0 (\vec{x}_1,w_1,\vec{x}_2,w_2)\,.
\end{align}
This computation shows that the second terms are also same and we proved that the boundary channel \eqref{bounddecosmall} and the ambient channel \eqref{ambient} are the same exactly.
In this proof we do not assume any specific form for $\delta V (r)$.
We can apply this proof to any case.

\section{Example: The Janus iCFT}
\label{sec4}

In this section, we consider the Janus iCFT as an example.
The coupling constant of the Janus iCFT jumps across the interface.
From the bulk point of view, this is because the dilaton field is not constant and has a non-trivial profile.
Regarding the dilaton as a source, we consider the 2-pt function of operators dual to axions.

The bulk dual of the Janus iCFT is a solution of type IIB supergravity.
The dilaton field $\phi$ of Janus has a non-trivial profile and depends only on the radial coordinate. The dilaton satisfies the equation of motion,
\begin{equation}
\partial_M (\sqrt{-g} g^{MN}\partial_N \phi )=0 \,.
\end{equation}
For a dilaton with only dependence on the radial coordinate $r$ this implies
\[
\partial _r \phi = \varepsilon \, \mathrm{e}^{-dA}
\]
where $\varepsilon$ an integration constant and will be assumed to be small when we consider perturbation theory. $\varepsilon$ is proportional to the jump in coupling constant in the dual iCFT. The corresponding correction to the metric is of order $\varepsilon^2$ and so, as in \eqref{small}, can be neglected to obtain the leading order correction to the 2-pt function.

The action of type IIB supergravity in the Einstein frame contains a coupling term between the axion field $a$ and the dilaton field,
\begin{equation}
S= \ldots + \frac{1}{2\kappa^2}\int \! \mathrm{d} ^{10} x  \sqrt{-g} \, \mathrm{e}^{-2\phi} g^{MN} \partial_M a \, \partial_N a \,.
\end{equation}
Since the dilaton field is not constant, the 2-pt function of the dimension 4 operator $\Tr F \wedge F$ dual to the axion field will be modified by the dilaton field.
Plugging a mode expansion of the axion field as in \eqref{sov}
\[
a= \sum_n \psi _n(r) \phi_{d,n} (y)
\]
into the equation of motion of the axion
\begin{equation}
\frac{1}{\sqrt{-g}}\partial_M (\sqrt{-g}g^{MN} \mathrm{e}^{-2\phi} \partial_N a)=\left( \mathrm{e}^{-2\phi } \mathcal{D}_r^2+\mathrm{e}^{-2A-2\phi }\partial_d^2+\partial_r \mathrm{e}^{-2\phi} \cdot \partial _r  \right) a=0 \,,
\end{equation}
it reduces to
\[
\left( \mathrm{e}^{-2\phi } \mathcal{D}_r^2+\mathrm{e}^{-2A-2\phi }m_n^2+\partial_r \mathrm{e}^{-2\phi} \cdot \partial _r  \right) a=0\,.
\]
Naively, if we expand the dilaton term in the above equation in $\varepsilon$ and regard the leading term of order $\varepsilon$ as $\delta V$, we may obtain leading correction terms
to modefunctions and energy.
But the problem is not so simple because this naive potential $\delta V$ contains a first derivative term about $r$ and the dilaton term appears in front of the energy term.
Thus, a more careful treatment is required to discuss perturbation theory in this case.

As noted in the previous footnote \ref{foot5}, to use standard quantum mechanical perturbation results we should first change the variable from $r$ to $z$ with
$\mathrm{d}r=\mathrm{e}^A \mathrm{d}z$.
Furthermore, we rescale the field as $\psi_n=\mathrm{e}^{-(d-1)A/2+\phi} \Psi_n$ to remove the first derivative term.
Then the problem reduces to standard quantum mechanics with an energy $E_n=m_n^2/2$, a kinetic term $-(1/2)\mathrm{d}^2/\mathrm{d}z^2$, a potential
\begin{equation}
V(z)=\frac{1}{2}\left[ \left( \frac{d-1}{2}\frac{\mathrm{d}A}{\mathrm{d}z}-\frac{\mathrm{d}\phi}{\mathrm{d}z}\right)^2+\frac{d-1}{2}\frac{\mathrm{d}^2A}{\mathrm{d}z^2}-\frac{\mathrm{d}^2\phi}{\mathrm{d}z^2}+M^2\mathrm{e}^{2A+2\phi}\right]
\end{equation}
(where we introduce a mass term though this term does not appear in the equation of motion of the axion) and a standard normalization\footnote{When the dilaton term is in front of the kinetic term, the original orthogonality relation is
\[
\int \! \mathrm{d} r\, \mathrm{e} ^{(d-2)A-2\phi} \psi_m \psi_n =\delta_{mn}\,.
\]
}
\begin{equation}
\int_{-\pi/2}^{\pi/2} \! \mathrm{d}z \, \Psi_m^0 (z)\Psi_n^0 (z)=\delta_{mn}\,.
\end{equation}

Let's return to the discussion of perturbation due to the dilaton profile.
Expanding the potential to leading order in $\varepsilon$, we obtain
\begin{equation}
\varepsilon \, \delta V (z)=-\frac{1}{2}\left[ (d-1) \partial_z A  \partial_z \phi + \partial_z ^2 \phi  \right]
=\frac{\varepsilon}{2} \sin z \cos^{d-1} z\,.
\end{equation}
Note that $\delta V (z)$ in Janus is an odd function and hence the leading correction to the eigenvalue vanishes.
Finally we obtain
\begin{equation}
\label{gammamn}
\gamma_{mn}=2\int_{-\pi/2}^{\pi/2}\! \mathrm{d}z \, \frac{\Psi_m^0(z)\delta V(z) \Psi_n^0(z) }{(m_n^0)^2-(m_m^0)^2} \,.
\end{equation}
The modefunctions are summarized in appendix \ref{modefunctions}.
If we introduce a new variable $x=\sin z$, the integrand reduces to products of
$x$, $1-x^2$ and associated Legendre polynomials.
Using the following two relations
\begin{align}
xP_\nu^\mu (x)&=\frac{(\nu- \mu +1)P_{\nu+1}^\mu(x)+(\nu+\mu)P_{\nu-1}^\mu (x)}{2\nu +1} \,,\\
\sqrt{1-x^2}P_\nu^{\mu}(x)&= \frac{P_{\nu+1}^{\mu +1}(x)-P_{\nu-1}^{\mu +1}(x)}{2\nu+1}
\end{align}
iteratively, the integral finally reduces to sum of products of two associated Legendre polynomials.
When $\Delta_n-d/2$ and $\Delta-d/2$ are both integers,
associated Legendre polynomials satisfy the orthogonality relation,
\begin{equation}
\int_{-1}^{1}\! \mathrm{d}x \, P_n^m(x)P_\ell^m(x)=
\begin{cases}
\frac{2(n+m)!}{(n-m)!(2n+1)}\delta_{n \ell} & \mbox{for } m\leq n \\
0 & \mbox{for } m>n
\end{cases}
\end{equation}
and we can compute $\gamma_{mn}$ explicitly.
The result are however complicated and not very illuminating, so we do not give the explicit expressions beyond \eqref{gammamn}.

\section{Conclusion and Discussion}
\label{sec5}

In this paper we discussed the conformal block expansion of 2-pt functions in general holographic bCFTs.

In section \ref{3.2}, we provided the decomposition of the 2-pt function in the boundary channel.
This was accomplished by decomposing the bulk-to-bulk propagator on the full $d+1$ dimensional geometry into the radial direction and AdS$_d$ space.
It was shown that conformal blocks in the boundary channel are given by bulk-to-bulk propagators on the AdS$_d$ slice.
We also obtained the leading correction of 2-pt functions by the perturbation around pure AdS and background fields.

We also confirmed that our conformal block expansion works in the case without boundary.
The summation of conformal blocks can be written as a 1-pt function of just the identity operator. This is an expected result because all 1-pt functions except that of the identity operator vanish without boundary, but reproducing this in the boundary channel expansion proved to be surprisingly tedious.

In section \ref{3.3}, we discussed the ambient channel.
We provided the leading correction due to conformal blocks in the ambient channel from first principles.
The contribution of a given conformal block contains products of two bulk-to-boundary propagators and one bulk-to-bulk propagator. They intersect at points on the geodesic as in the probe brane case \cite{Rastelli:2017ecj}. The remaining point connected to its bulk-to-bulk propagator couples to a source term.
When the source term is a delta function in the radial direction, our decomposition reduces to that of \cite{Rastelli:2017ecj}.
We also proved the equivalence between the two decompositions, boundary channel and ambient channel, in section \ref{3.4}.

In section \ref{sec4}, we considered $d$ dimensional Janus solutions as an example.
Since the $d=4$ Janus solutions is constructed from type IIB supergravity, the dual CFT is known explicitly. So, the Janus geometry is a good example.
In Janus, the source is an odd function with respect to the radial coordinate $r$.
Hence, the conformal dimension is not affected by the source and only $\gamma_{mn}$ is non-trivial.
We explicitly computed the potential and obtained $\gamma_{mn}$ as integrals over Legendre polynomials .
Our prescription can be easily generalized to other cases.

We would like to comment on the relation between our paper and \cite{Rastelli:2017ecj}.
Ref. \cite{Rastelli:2017ecj} only addressed a situation where a defect is a probe brane at $r=0$.
Our paper considers  more general boundary or defect CFTs. In addition, we were able to derive our prescription and so, in principle, can easily generalize it to higher orders. Most notably, our boundary channel decomposition into blocks, \eqref{bounddeco} and \eqref{coeffs}, is exact.
As we saw sections \ref{3.2} and \ref{3.3}, our results \eqref{bounddecosmall} and \eqref{ambient} include those in \cite{Rastelli:2017ecj}.

\section*{Acknowledgements}
We would like to thank Christopher Herzog for useful correspondence.
The work of AK was partly supported by the US Department of Energy under grant number DE-SC0011637. The work of YS is supported by a JSPS fellowship.

\appendix

\section{AdS$_{d+1}$ Modefunctions}
\label{modefunctions}

In this appendix, we summarize eigenfunctions of the differential operator $\mathcal{D}_r^2+\mathrm{e}^{-2A}m_n^2-M^2$ when $\mathrm{e}^A=\cosh(r)$, that is for unperturbed AdS$_{d+1}$, and derive some useful formulas among them. We use $m_n^2=\Delta_n(\Delta_n-(d-1))$ and $M^2=\Delta (\Delta-d)$ to label the eigenvalue of the mode and the bulk mass in terms of the operator dimension appearing in the BOPE and the dimension of the ambient space operator.
The eigenfunctions are given by
\begin{align}
\psi_n(r)&=\frac{C_n}{2^{\Delta}\cosh ^\Delta r}{}_2F_1\left(\frac{\Delta-\Delta_n}{2}, \frac{\Delta+\Delta_n-d+1}{2} ,\Delta -\frac{d}{2}+1 ,\frac{1}{\cosh ^2 r} \right) \\
&=\frac{C_n\Gamma (\Delta-d/2+1)}{2^{d/2} \cosh^{\frac{d}{2}}r}P_{-1+\frac{d}{2}-\Delta_n}^{\frac{d}{2}-\Delta}(\tanh r)
\end{align}
where $C_n$ is normalization constant.
These two different expressions can be shown to be the same by using the following identities,
\begin{equation}
{}_2F_1\left(2\alpha,2\beta, \alpha+\beta+\frac{1}{2},z\right)={}_2F_1\left(\alpha,\beta, \alpha+\beta+\frac{1}{2},4z(1-z)\right)
\end{equation}
and
\begin{equation}
{}_2F_1(\alpha,\beta, \gamma,z)=(1-z)^{\gamma -\alpha -\beta}{}_2F_1(\gamma-\alpha,\gamma-\beta, \gamma,z) \,.
\end{equation}
From the boundary condition that the eigenfunctions are normalizable at large $r$, $\Delta_n$ are determined as
\begin{equation}
\Delta_n=\Delta+n \,.
\end{equation}
This is the expected relation given the interpretation of the BOPE as a Taylor expansion in the no-brane case.

The normalization constant can be determined explicitly.
Introducing a new variable $x=1/\cosh ^2 r$ and using a functional identity
\begin{equation}
{}_2F_1 \left(a,b,a+b+\frac{1}{2} ,z \right)^2={}_3F_2 \left(2a,2b, a+b ,2a+2b,a+b+\frac{1}{2},z\right)\,, 
\end{equation}
we obtain
\begin{align}
\frac{1}{C_n^2}&= \int _0^1 \! \mathrm{d}x \, x^{a+b-1/2} (1-x)^{-1/2}{}_3F_2 \left(2a,2b, a+b ,2a+2b,a+b+\frac{1}{2},x\right) \notag \\
&= \frac{\sqrt{\pi}\Gamma (a+b+1/2) }{2^{2\Delta}\Gamma (a+b+1)} {}_3F_2 \left(2a,2b, a+b ,a+b+1,2a+2b,1\right)
\end{align}
where $a=(\Delta-\Delta_n)/2=-n/2$ and $b=(\Delta+\Delta_n-d+1)/2$ are introduced to simplify the expressions.
Furthermore, the hypergeometric ${}_3F_2$ at $x=1$ is evaluated as 
\begin{equation}
{}_3F_2 (-n,\alpha ,\beta, \gamma ,1+\alpha +\beta -\gamma -n, 1 )=
\frac{(\gamma -\alpha)_n (\gamma -\beta)_n }{(\gamma)_n (\gamma-\alpha-\beta)_n}
\end{equation}
for $n \in \mathbb{N}$. 
After some computations, the normalization constant is analytically determined as
\begin{equation}
C_n^2=\frac{2^{d-1} \Gamma(\Delta_n+\Delta-d+1)\,  (2\Delta_n-d+1)}{n! \, \Gamma(\Delta -d/2+1)^2}\,.
\end{equation}

\bibliographystyle{JHEP}
\bibliography{boundarygeodesicwitten}

\end{document}